\documentclass[fleqn,usenatbib]{mnras}

\usepackage{newtxtext,newtxmath}

\usepackage[T1]{fontenc}

\DeclareRobustCommand{\VAN}[3]{#2}
\let\VANthebibliography\thebibliography
\def\thebibliography{\DeclareRobustCommand{\VAN}[3]{##3}\VANthebibliography}

\usepackage{graphicx}	
\usepackage{amsmath}	

\newcommand{\mps}{\; {\rm m\;s^{-1}}}

\newcommand{\btt}{ }

\def \tdrag {\tau_\mathrm{drag}}

\title[Atmospheric circulation of brown dwarfs and directly imaged exoplanets]{Atmospheric circulation of brown dwarfs and directly imaged exoplanets driven by cloud radiative feedback: global and equatorial dynamics}

\author[Tan \& Showman]{
Xianyu Tan$^{1}$\thanks{E-mail: xianyu.tan@physics.ox.ac.uk}
and Adam P. Showman$^{2,3,\dagger}$
\\
$^{1}$Atmospheric Oceanic  and Planetary Physics, Department of Physics, University of Oxford, OX1 3PU, UK\\
$^{2}$Lunar and Planetary Laboratory, University of Arizona, 1629 University Boulevard, Tucson, AZ 85721, USA \\
$^{3}$Department of Atmospheric and Oceanic Sciences, Peking University, Beijing, People’s Republic of China\\
$^{\dagger}{\rm Deceased.}$
}

\date{Accepted XXX. Received YYY; in original form ZZZ}

\pubyear{2021}

\begin{document}
\label{firstpage}
\pagerange{\pageref{firstpage}--\pageref{lastpage}}
\maketitle

\begin{abstract}
Brown dwarfs, planetary-mass objects and directly imaged giant planets exhibit significant observational evidence for active atmospheric circulation, raising critical questions about mechanisms driving the circulation, its fundamental nature and time variability. {\btt Our previous work has demonstrated the crucial role of cloud radiative feedback on driving a vigorous atmospheric circulation using local models that assume a Cartesian geometry and constant Coriolis parameters.} In this study, we extend the models to a global geometry  and explore properties of the global dynamics. {\btt We show that, under relatively strong dissipation in the bottom layers of the model, horizontally isotropic vortices are prevalent at mid-to-high latitudes while large-scale zonally propagating waves  are dominant at low latitudes  near the observable layers. } The equatorial waves have both eastward and westward phase speeds, and the eastward components with typical velocities of a few hundred $\mps$ usually dominate the equatorial time variability. Lightcurves of the global simulations show variability with amplitudes from 0.5 percent to a few percent depending on the rotation period and viewing angle. {\btt The time evolution of simulated lightcurves is} critically affected by the equatorial waves, showing wave beating effects and differences in the lightcurve periodicity to the intrinsic rotation period. The vertical extent of clouds is the largest at the equator and decreases poleward due to the increasing influence of rotation with increasing latitude. {\btt Under weaker dissipation in the bottom layers, strong and broad zonal jets develop and modify wave propagation and lightcurve variability.}   Our modeling results help to qualitatively explain several features of observations of brown dwarfs and directly imaged giant planets, including  puzzling time evolution of lightcurves, a slightly shorter period of variability in IR than in radio wavelengths, and the viewing angle dependence of variability amplitude and IR colors.
\end{abstract}




\section{Introduction}
\label{ch.intro}

Brown dwarfs (BDs) are objects with masses intermediate between stars and giant planets, but are analogous to giant planets in temperature, composition and size \citep{burrows2001}. Isolated BDs are easier to observe than exoplanets orbiting bright stars, making them ideal targets to investigate physical, chemical and dynamical processes in the context of planetary atmospheres \citep{apai2017}. Growing observations of BDs in the past decade have revealed that active atmospheric circulation is  common among BDs.
Here we summarize several key types of observations that set direct  constraints on the large-scale circulation:
\begin{itemize}
    \item {\it Broadband lightcurve time variability} at infrared (IR) wavelengths is commonly observed for field BDs   (e.g., \citealp{gelino2002,clarke2008, artigau2009, gillon2013,buenzli2014,wilson2014, metchev2015,leggett2016,MilesPaez2017, apai2017,manjavacas2017,vos2019,eriksson2019,bowler2020,Hitchcock2020,vos2020}, see also recent reviews by \citealp{biller2017} and \citealp{artigau2018}). Most of them are likely caused by the rotation of large-scale surface inhomogeneities related to different  cloud opacity  and/or temperature that move in and out of  view as the objects rotate. The shapes of some lightcurves are irregular and change over timescales comparable to or slightly longer than  rotational timescales, indicating rapid evolution of the large-scale spatial patchiness in some cases. \cite{apai2017} summarized a few types of irregularities of the lightcurve variability and suggested a possible solution by invoking differential zonal propagation of waves.
    
    \item {\it Multi-wavelength near-IR lightcurve  variability} provides additional information on the vertical structure of the surface patchiness  because different wavelengths probe different atmospheric pressures \citep{buenzli2012,radigan2012,biller2013,apai2013,yang2015,yang2016,lew2016,zhou2018,zhou2020,lew2020}. The amplitude of the variability changes with  wavelength, showing smaller amplitude near the water absorption bands and larger amplitude near spectral windows. These are typically consistent with the scenario that the surface patchiness results from a combination of different cloud thicknesses rather than a combination of a single cloud type and a completely clear-sky region \citep{buenzli2012,apai2013,yang2015,lew2016,lew2020}. Alternative scenarios invoking variation of temperature in clear-sky atmospheres have also been proposed \citep{robinson2014,morley2014b, tremblin2020}.    The phases of lightcurve rotational variations can be different at different wavelengths, and the extent of this difference varies with spectral types \citep{yang2016}. 
    
    \item {\it Viewing-angle dependence of variability and near-IR color} of field BDs has been suggested by \cite{vos2017}, \cite{vos2018} and \cite{vos2020}, who showed that  BDs viewed more equator-on tend to have higher variability amplitudes and redder near-IR colors than those viewed more pole-on. These results indicate the possible presence of systematic equator-to-pole differences of cloud properties, temperature and/or chemical composition.
    
    \item {\it Doppler imaging} could provide perhaps the most direct constraints on the presence of large-scale surface patchiness and their morphology. This has been done for  Luhman 16B (the spectral type T1  component in the BD binary system Lhuman 16AB) and yielded surface flux differences between the patchiness of a few hundred K in terms of effective temperature \citep{crossfield2014}.
    
    \item {\it Simultaneous tracking of near-IR and radio variability}, in which the former measures the  period at which atmospheric features rotate in and out of the view and the latter likely reflects the  rotation period of the interior, could track possible differential rotation between the atmosphere and the interior. Recently, \cite{allers2020} applied this technique to a nearby T6 brown dwarf  and showed that the period of the near-IR variability is slightly shorter than that of the radio emission, suggesting that the dominant atmospheric features travel eastward relative to the interior with a zonal speed of a few hundred$\mps$. 
    
    \item {\it Net near-IR polarization} could be  due to scattering by cloud particles  together with either the presence of oblateness due to fast rotation or  cloud inhomogeneity (e.g., \citealp{sengupta2009,marley2011}).  Recently, \cite{millar2020} unambiguously detected net near-IR polarization as well as its time variability on Luhman 16B. The time-mean polarization indicates a latitudinally dependent cloud structure, and its  time variation indicates longitudinal cloud patchiness on top of the zonal cloud structure.
\end{itemize}


Directly imaged extrasolar giant planets (EGPs)  are mostly young, hot and relatively distant from their host stars such that they receive negligible external stellar irradiation compared to their interior heat flux. Their spectrum and near-IR colors show similarities to the dusty field BDs (e.g., \citealp{currie2011,  barman2011a, marley2012, rajan2015,chauvin2017, stolker2020}). Near-IR lightcurve variability has  been observed on directly imaged EGPs and   planetary-mass, free-floating  objects \citep{biller2015,zhou2016,vos2018,biller2018,schneider2018,manjavacas2019,miles2019}. From a meteorological point of view, directly imaged EGPs resemble low-gravity versions of BDs and  fall into the same category as field BDs in terms of atmospheric dynamical regime.

These observations motivate the investigation of global atmospheric dynamics of BDs and directly imaged EGPs, their fundamental properties and effects on cloud formation and chemistry.  There have been several studies investigating atmospheric dynamics appropriate for these objects (\citealp{freytag2010,zhang&showman2014,tan2017,showman2019,tan2020bd}, and see  recent reviews by \citealp{showman2020} and \citealp{zhang2020}). Cloud radiative feedback has been proposed as a robust and novel mechanism to drive spontaneous atmospheric variability and dynamics in these atmospheres \citep{tan2019,tan2020bd}. This mechanism works as  follows: imagine two adjacent atmospheric columns, one of which has a thick cloud layer and the other has a relatively thin cloud layer. The two columns experience different radiative heating and cooling rates due to the different cloud opacity, which then generate  isobaric temperature differences. This drives an overturning flow that maintains the  patchy clouds against settling, potentially sustaining the circulation.   In a previous work, we confirmed   the viability of this mechanism and explored dynamical properties of the circulation  using local three-dimensional models that assume a Cartesian geometry and a constant Coriolis parameter $f$ across the domain \citep{tan2020bd}.  We demonstrated the importance of rotation  in regulating the  large-scale atmospheric dynamics and cloud formation, showing that the  typical horizontal length scales of vortices and cloud patterns are inversely proportional to the Coriolis parameter $f$ if other parameters are held fixed, and that the mean cloud vertical extent decreases with decreasing rotation period.  

The latitudinal variation of the Coriolis parameter in  global geometry, the so-called $\beta$ effect where $\beta=df/dy$ and $y$ is distance increasing northward,  introduces additional dynamical behaviors compared to those in the $f$-plane models \citep{tan2020bd}. For example, the $\beta$ effect can introduce horizontal anisotropy in the turbulence \citep{rhines1975,vallis1993}  as well as large-scale atmospheric waves \citep{holton2012}, both of which play crucial  roles in the global circulation of giant planets (see reviews by  \citealp{ingersoll2004,vasavada2005, del2009,showman2018review}). In addition, the regional models in \cite{tan2020bd} occupy only a limited fraction of the surface area of  BDs and EGPs,  and therefore qualitative comparisons between model outputs and observed lightcurve variability were lacking. 

In this study,  we  extend the modeling framework of \cite{tan2020bd} to a global geometry to investigate the global atmospheric circulation driven by the cloud radiative feedback and the resulting lightcurve variability.  We show that when the bottom frictional dissipation is relatively strong, horizontally isotropic storms and turbulence are prevalent at mid-to-high latitudes while zonally propagating waves are present at low latitudes near the observable layer. The differential propagation of equatorially trapped waves induces short-term evolution of simulated lightcurves, analogous to the wave beating effects described in \cite{apai2017}. Eastward propagating equatorial Kelvin waves are sometimes dominant, causing a slightly shorter rotation period of the atmospheric features relative to the underlying planetary rotation period, which agrees well with observational results by \cite{allers2020}.  When the bottom dissipation is weak, strong and broad zonal jets develop and modify wave propagation and lightcurve variability. We find systematic equator-to-pole differences of clouds and temperature structures due to  the latitudinal variation of the Coriolis parameter  $f$,  supporting recent observations by \cite{vos2017} and \cite{vos2020}. Large-scale equatorial disturbances may help to explain the nature of longitudinal variation  in Doppler mapping \citep{crossfield2014}  and time-varying polarization \citep{millar2020} of Luhman 16B. 

The paper is organized as follows. Section \ref{ch.model} briefly  introduces the global numerical model. Section \ref{ch.result} describes results of models with different rotation periods and drag timescales as well as their resulting lightcurve variability. In Section \ref{ch.conclusion} we discuss  implications of the results and  draw conclusions.

\section{Model}
\label{ch.model}
The general circulation model (GCM) used in this study is fully described  in \cite{tan2020bd}, and here we apply the GCM to a global domain. Key elements of the model are summarized below. We solve the  standard 3D hydrostatic primitive equations using an atmospheric GCM, the MITgcm \citep{adcroft2004}. Two tracer equations which represent the mass mixing ratio of condensable vapor and clouds are integrated simultaneously. We assume the ideal gas law  for the equation of state. {\btt The deep layers in our models have reached the convective zone which is typically at pressures larger than a few bars for typical L and T dwarfs \citep{burrows2001}. Effects of rapid   convective mixing  on both entropy and tracers are parameterized using a simple convective adjustment scheme that instantaneously homogenizes entropy and tracers in the convectively unstable regions. }

{\btt A Rayleigh drag is applied to the horizontal winds  at pressures higher than 5 bars to crudely represent interactions between the modeled atmosphere and the quiescent interior where large-scale flows are likely retarded due to significant magnetohydrodynamic dissipation.  Rapid convective mixing of specific entropy and fast rotation may lead to the Taylor–Proudman effect that could be efficient in slowing down large-scale winds in the shallow outer layer (see the detailed argument in \citealp{schneider2009}). Because of the much hotter interior of L and T dwarfs and likely strong magnetic fields of BDs ($\sim$kG, \citealp{kao2016,kao2018}), the ``quiescent" region inside BDs is expected to extend to a  much shallower depth than that of Jupiter. Therefore the drag is applied uniformly in the horizontal direction. The form and strength of the drag are highly uncertain because of the unknown nature of interactions between the interior and the  shallow outer layer. Yet no theoretical study systematically investigate how such interactions should be parameterized in GCMs of gaseous planets. Nevertheless, this simple drag  serves to  dissipate flows in our bottom model layers, and  similar  drag schemes have been used in previous studies of Jupiter \citep{schneider2009} and hot Jupiters (e.g., \citealp{liu2013,tan2019uhj,carone2020}). }

{\btt The drag is the strongest at the bottom boundary and is characterized by a drag timescale $\tdrag$, then it  linearly decreases with decreasing pressure until it reaches zero at 5 bars. There is no  drag at pressures lower than 5 bars.  As in \cite{tan2020bd}, we set a relatively strong drag timescale $\tdrag=10^5$ s for the major suit of models. We have tested models with stronger drags ($\tdrag=10^4$ and $10^3$ s), and they are  quantitatively  similar to those with $\tdrag=10^5$ s.  Therefore, $\tdrag=10^5$ s is chosen to represent dynamics in the ``strong-drag" regime. For $\tdrag$ sufficiently larger than $10^5$ s, dynamics is in the weak-drag regime. The drag is said to be ``strong" in a practical sense that the resulting zonal-mean zonal flows near the bottom layer are rather weak (with velocities much smaller than $100\mps$) based on our modeling results discussed in Section \ref{ch.drag}, whereas the drag is said to be ``weak" when the resulting zonal flows near the bottom layer are strong (with velocities much greater than $100\mps$). Using  scaling argument in the convective layers, \cite{showman&kaspi2013} argue that the characteristic speeds of large-scale  zonal flows near the top of convective zone  range from only a few to tens of $\mps$ depending on the number of jets.    We therefore expect that the strong-drag regime might be  appropriate for BDs and hot directly imaged EGPs.    Nevertheless, in section \ref{ch.drag}, we show results with weaker drags of $\tdrag=10^6$ and $10^7$ s to explore possible circulation patterns for two reasons. First, it is theoretically motivated to understand dynamics in different regimes in spite of its applicability in realistic situations. Second, recent gravity measurements of Jupiter and Saturn suggest that significant meridional density gradients (and therefore the vertical wind shears associated with the zonal jets) exist deep in the convective zone of Jupiter and Saturn \citep{kaspi2018,guillot2018,iess2019}. This indicates that  strong organized zonal flows near the top of convective zone    remain possible for BDs.  }

Atmospheric motions are driven by the horizontal pressure gradient that is rooted from horizontal differential radiative heating and cooling. In calculating the radiative flux, we assume a gray atmosphere with a single broad thermal band for simplicity and computational efficiency. {\btt The gas opacity in our model atmosphere is $\kappa_{\rm{gas}} = \max(\kappa_{\rm{R, g}}, \kappa_{\rm{min}})$, where $\kappa_{\rm{R, g}}$ is the Rosseland-mean opacity  from \cite{freedman2014} assuming solar composition.  The Rosseland-mean opacity gives a good estimation of radiative flux in the optically thick limit. The atmosphere above about 1 bar is optically thin by the gas opacity alone, and  there is no good choice {\it a priori} for the opacity in the grey approximation.   A   minimal opacity $\kappa_{\rm{min}}$ is then imposed in the gas opacity. We assume $\kappa_{\rm min}=10^{-3}\;{\rm m^2\;kg^{-1}}$, which is consistent with that used in semi-grey models in \cite{guillot2010} for hot giant planets.} The total atmospheric opacity $\kappa$ is simply the sum of the gas and cloud opacities $\kappa = \kappa_{\rm{gas}} + \kappa_c$,  the latter of which is determined by the cloud mixing ratio as a function of time and location.

The cloud formation scheme is the same as {\btt that in} \cite{tan2020bd}. Cloud forms when the mixing ratio of condensable vapor exceeds the prescribed saturation mixing ratio $q_s$. Otherwise, evaporation occurs when the vapor mixing ratio is less than $q_s$.  The saturation mixing ratio $q_s$ is assumed to depend on pressure alone: $q_s$ is $q_{\rm{deep}}$ at a condensation pressure $p_{\rm cond}$ and then  rapidly decreases when pressure is less than $p_{\rm cond}$. At pressures larger than $p_{\rm cond}$, $q_s$  is assumed to be arbitrarily large  such that no condensation would occur. {\btt The condensation pressure $p_{\rm cond}$ is assumed to be 0.5 bar, roughly consistent with conditions of typical mid-L dwarfs in which clouds form at and above $\sim 1$ bar (e.g., \citealp{tsuji2002,burrows2006}).} The  deep mixing ratio $q_{\rm{deep}}$ is assumed to be $2\times10^{-4} ~ \rm{kg/kg}$. At pressures higher than 5 bars, the vapor field is relaxed towards $q_{\rm deep}$ over a timescale of $10^3$ s. {\btt  The assumed  $q_{\rm deep}$ in this study is less than that of silicate vapor in atmospheres with solar abundance. This value is tuned to generate cloud opacity that greatly exceeds the gas opacity, but not to trigger convection {\it within} the cloud-forming region. Convection within the cloud forming region complicates the dynamics as it would introduce instability that does not totally rely on resolved large-scale flow \citep{tan2019}.  We aim at exploring large-scale cloud-driven dynamics in a cleanest possible context, therefore we do not chose a realistic $q_{\rm deep}$. } 

We assume a constant cloud particle number  per dry air mass  $\mathcal{N}_c$ (in unit of $\rm{kg^{-1}}$) throughout the atmosphere.   Cloud particles are assumed to be spherical and have a single size locally in each grid cell, and the particle size is then determined via a  given $\mathcal{N}_c$ and time- and location-dependent amount of condensate (using the scheme of \citealp{tan2020bd}). {\btt We assume  $\mathcal{N}_c=10^{11} ~\rm{kg^{-1}}$, which results in typical cloud particle size around 0.5 $\mu$m given the assumed $q_{\rm deep}$, consistent with that expected for L dwarfs \citep{saumon2008,burningham2017}.} Optical properties of cloud particles, including the extinction coefficient,  scattering coefficient and asymmetry parameter are averaged  using the Rosseland-mean definition. Tables containing these parameters as functions of temperature and pressure  are  pre-calculated  for the GCM  to read in during the integration.  We use enstatite (${\rm MgSiO_3}$) to represent properties of the cloud species, including a density $\rho_c = 3190 \;{\rm kg\;m^{-3}}$ and the refractive index. {\btt Enstatite is one of the most representative cloud species in atmospheres of L and early T dwarfs, and is appropriate for our atmospheric conditions \citep{ackerman2001,saumon2008,marley2015}. However, the choice of cloud species is not critical in this study as long as the cloud opacity exceeds the gas opacity.   }

The global models in this study have different horizontal resolution depending on the rotation period---the shorter the rotation period, the smaller the deformation radius and thus higher resolution is needed. Our horizontal resolution of the cubed-sphere grid is C96 (equivalent to 384$\times$192 grid points in longitude and latitude), C128 (512$\times$256), C192 (768$\times$384) and C256 (1024$\times$512) for rotation period of 20, 10, 5 and 2.5 hours, respectively. These horizontal resolutions are among the highest  published so far for exoplanet and brown dwarf global models\footnote{If the horizontal resolution is insufficient to fully resolve dynamics within the deformation radius of local regions (especially near the polar regions), storms will cease to exist and there would not be dynamics in these regions.}. The radius of the object is assumed to be $7\times 10^7$ m, similar to the Jovian radius $R_J$. {\btt Although radius of real BDs varies from more than $2R_J$ to slightly less than $R_J$ depending on their masses and ages, our choice of radius should be representative for field BDs.}   A standard fourth-order Shapiro filter is applied in the horizontal velocity and temperature fields to maintain numerical stability \citep{shapiro1970}. The pressure domain is between 10 bars and $10^{-3}$ bar, which is discretized evenly into 55 layers in log pressure. {\btt The model pressure domain is deep enough to reach the convective zone, where we argue that rapid convective mixing leads to small horizontal temperature gradient. The upper boundary is high enough for  models to safely capture dynamics associated with clouds.   } The temperature at the bottom boundary (10 bars) is fixed at 2600 K, resulting in atmospheric temperatures comparable to those of L dwarfs.  We adopt physical parameters relevant for BDs, including the specific heat $c_p = 1.3\times 10^4 ~\rm{Jkg^{-1}K^{-1}}$, the specific gas constant $R= 3714 ~\rm{Jkg^{-1}K^{-1}}$, and a surface gravity $g=1000 ~\rm{ms^{-2}}$. {\btt  Models with $\tdrag=10^5$ s equilibrated after $\sim100$ simulation days. After equilibration, we  integrated them for  additional 300 days, and  the time-mean statistical results were obtained with outputs of the last 200 days. Models with $\tdrag=10^6$ and $10^7$ s equilibrated after $\sim200$ and $\sim1000$ simulation days, respectively, and their statistical results were obtained similarly to those with  $\tdrag=10^5$ s.  }

\section{Results}
\label{ch.result}

\subsection{Results with varying rotation and  strong bottom drag}

\subsubsection{Temperature and cloud patterns}
\begin{figure*}
	\includegraphics[width=1.9\columnwidth]{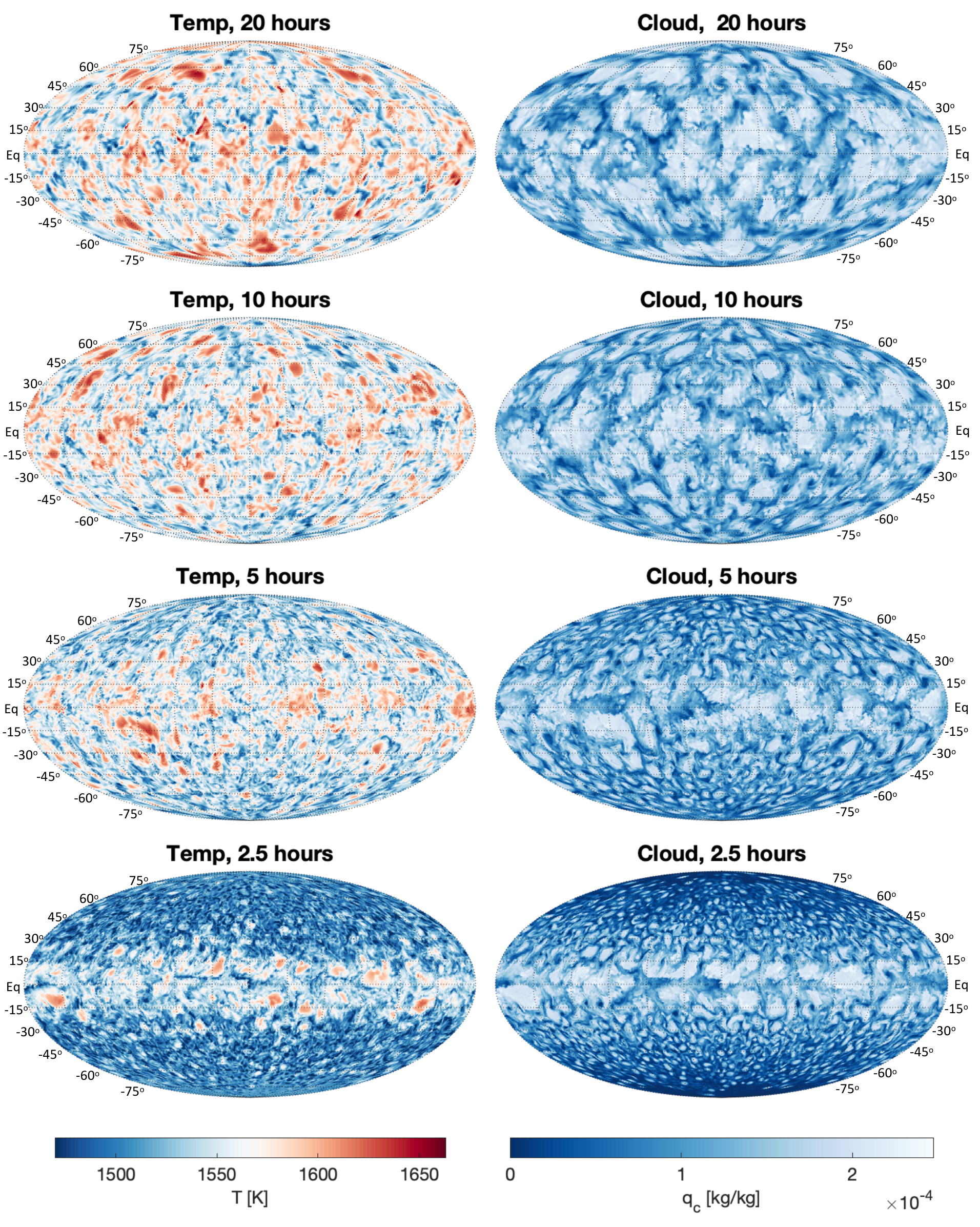}
    \caption{Instantaneous global temperature maps at 0.23 bar on the left column and the corresponding horizontal cloud mixing ratio at 0.23 bar on the right column. These results are from models with four rotation periods of 20, 10, 5 and 2.5 hours (from the top to the bottom row) and with a  drag timescale $\tau_{\rm{drag}}=10^5$ s at the bottom boundary. Horizontal thin dotted lines in the maps are constant-latitude lines with a spacing of $15^{\circ}$, and vertical thin dotted lines are constant-longitude lines with a spacing of $30^{\circ}$  }
\label{fig.global_tq}
\end{figure*}

\begin{figure*}
	\includegraphics[width=1.9\columnwidth]{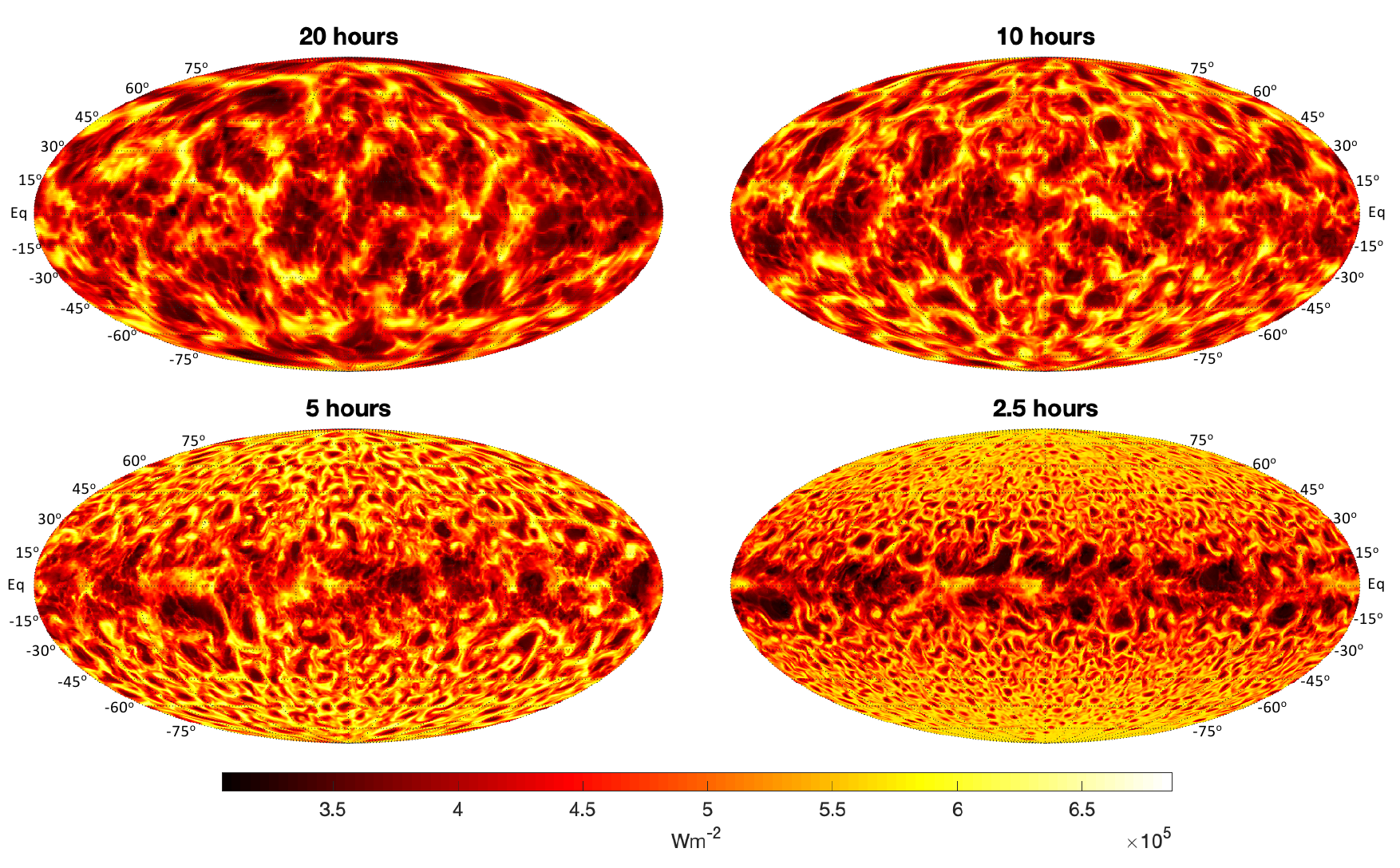}
    \caption{Instantaneous global  map of outgoing top-of-atmosphere thermal flux corresponding to those shown in Figure \ref{fig.global_tq}.  }
\label{fig.global_flux}
\end{figure*}

Results of global models with rotation periods of 20, 10, 5 and 2.5 hours and with a  drag timescale  $\tau_{\rm{drag}}=10^5$ s (note that the frictional drag is applied only at pressures greater than 5 bars) are shown in Figure \ref{fig.global_tq}.
Figure \ref{fig.global_flux} shows the corresponding outgoing top-of-atmosphere thermal flux. These results are obtained after the models reach statistical equilibrium. {\btt Pressures of thermally radiating levels are sensitive to cloud top pressures---the radiating pressure could be lower than 0.05 bar in cloudy regions and could be about 3 bars in clear-sky regions. 
Dynamics  at 0.23 bar is representative for layers  near and slightly below the cloud forming region. Quantities at slightly higher or lower pressures are qualitatively very similar to those at 0.23 bar. }

There is a lack of strong zonal jets, in the sense that the zonal-mean zonal winds are much weaker than the eddy velocities, in all models shown in Figure \ref{fig.global_tq} and \ref{fig.global_flux} due to dissipation of kinetic energy by the strong bottom drag. The circulation in the cloud forming region is  dominated by active vortices, turbulence and waves.    At 0.23 bar, typical isobaric temperature differences are above 100 K, and local horizontal wind speeds can exceed 1000$\mps$.    Mid-to-high latitudes are filled with  cyclones\footnote{Cyclones have relative vorticity the same sign as the local planetary rotation, whereas anticyclones have the opposite sign of relative vorticity.}  which are less cloudy than their surroundings, and anticyclones which are associated with thick  clouds. {\btt In the cloudy areas, less thermal radiation escapes to space due to large cloud opacity, and this cloud greenhouse effect warms up the lower cloud forming region. In the cloud-free areas,  more radiation can escape to space from the hotter, deeper region due to the lack of cloud opacity, efficiently cooling down the atmospheric column. This patchy-cloud configuration is self-maintained by the circulation driven by clouds themselves \citep{tan2020bd}. Positive buoyancy is generated in the cloudy regions, and it maintains the cloudiness against cloud gravitational settling by transport of condensable vapor upward to the condensation level; whereas negative buoyancy in the cloud-free regions generates downwelling that advects cloud-free air from above, maintaining the IR cooling. In rapidly rotating conditions, the tendency of geostrophy implies that the warm, cloudy regions are anticyclonic and cool, cloud-free regions are cyclonic.    }  The spatial patterns of the outgoing top-of-atmosphere thermal flux are highly correlated with that of  cloud patchiness. Other than the vortices,  turbulence and transient waves with smaller horizontal length scales and higher oscillation frequencies are also present. The basic findings from these global models agree well with those of box models with a fixed Coriolis parameter $f$ across the domain \citep{tan2020bd}.

At a given  rotation period, typical sizes of vortices are generally larger at lower latitudes and smaller  at higher latitudes. This is more prominent in rapidly rotating models  with rotation periods of 5 and 2.5 hours. The overall sizes of vortices are larger with longer rotation period. This is because the typical sizes of vortices driven by cloud radiative feedback  are close to the Rossby deformation radius $L_d=c_g/|f|$ \citep{tan2020bd}, where $c_g$ is the phase speed of gravity waves, $f=2\Omega\sin\phi$, $\Omega$ is planetary rotation rate and $\phi$ is latitude. The dependence of $L_d$ on $\Omega$ and $\phi$ naturally leads to the spatial and rotation-period dependence of vortex sizes.  Individual vortices are short lived and experiences merging, disruption, shearing and straining over typical timescales of tens of hours. There is no systematic migration of individual vortices after they form. This is likely because the  atmosphere is filled with vortices, and  individual vortex is significantly disrupted by adjacent vortices before they can migrate over a long distance.

\begin{figure*}
	\includegraphics[width=2\columnwidth]{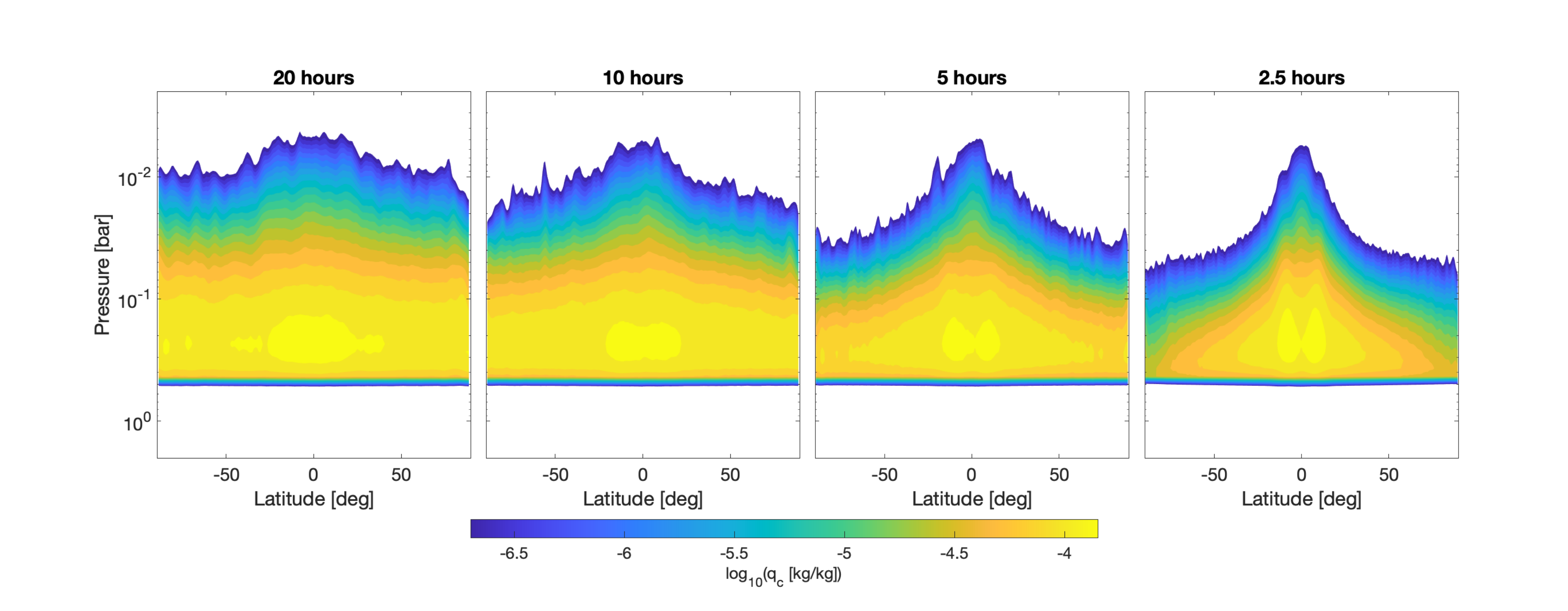}
    \caption{Time- and zonal-average cloud mass mixing ratio as a function of latitude and pressure for models with  four rotation periods of 20, 10, 5 and 2.5 hours (from the left to right) and with a  drag timescale $\tau_{\rm{drag}}=10^5$ s at the bottom. The white regions have $\log_{10} q_c < -7$.}
\label{fig.tracer2zonalav}
\end{figure*}

\begin{figure}
	\includegraphics[width=1\columnwidth]{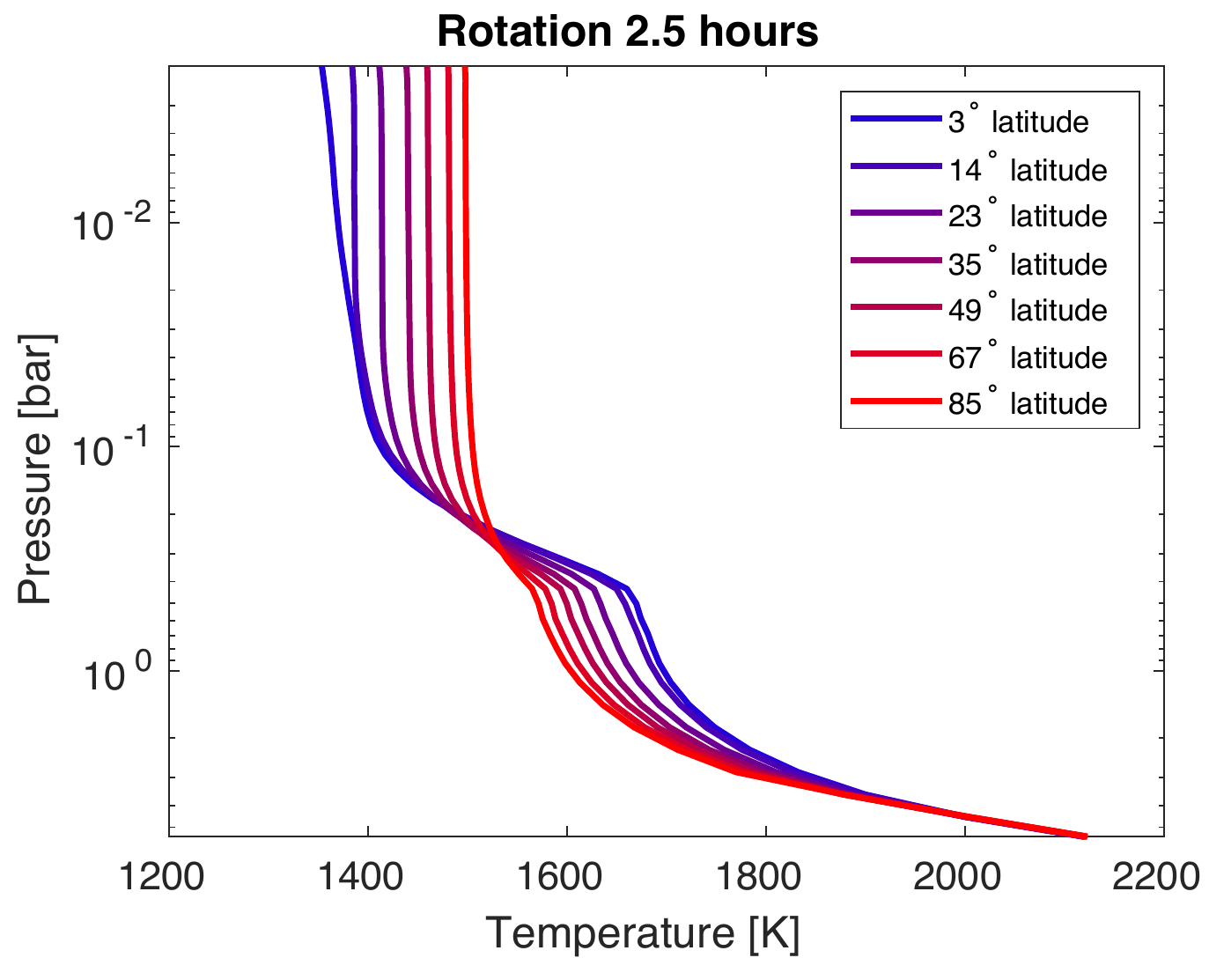}
    \caption{Time- and zonal-average temperature-pressure profiles at selected latitudes from the model with a rotation period of 2.5 hours and with a  drag timescale $\tau_{\rm{drag}}=10^5$ s at the bottom.}
\label{fig.zonalmeantp}
\end{figure}

\begin{figure*}
	\includegraphics[width=2\columnwidth]{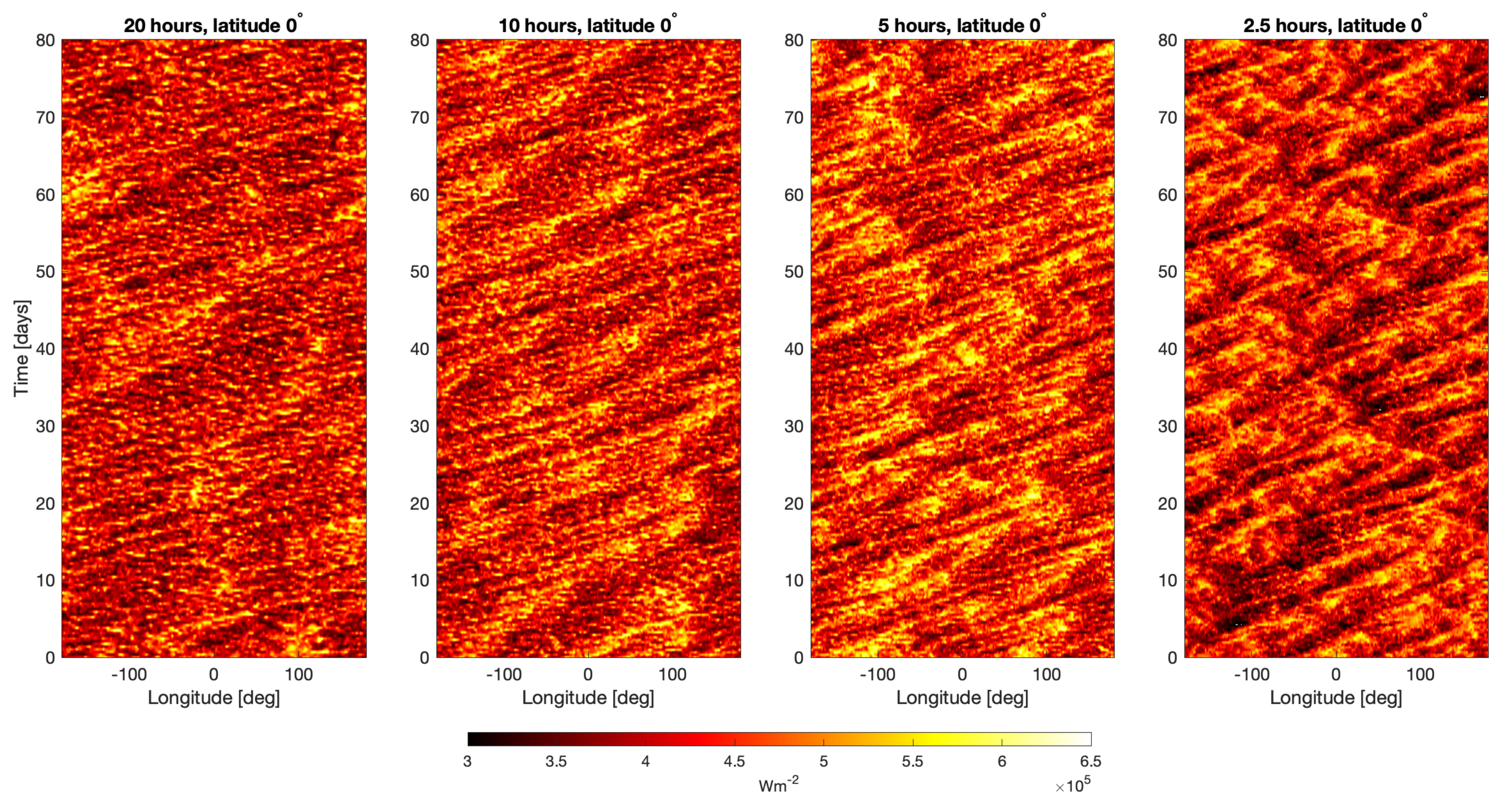}
    \caption{Hovm$\ddot{\rm o}$ller diagrams (longitude-time sequence) showing the time evolution of outgoing thermal flux at the equator for models with four rotation periods of 20, 10, 5 and 2.5 hours (from the left to right) and  a  drag timescale $\tau_{\rm{drag}}=10^5$ s. {\btt Fluxes shown in these panels are sampled every 0.3 day.}   }
\label{fig.equatortime}
\end{figure*}

As shown by \cite{tan2020bd}, greater $|f|$ leads to overall thinner clouds. Qualitatively,  stronger rotation leads to a larger degree of geostrophic balance in the flows which exerts  greater suppression to the vertical velocity, making the flows less efficient to vertically transport tracers against cloud gravitational settling.   Therefore, the variation of the Coriolis parameter $f$ from the equator  to the poles as well as the different rotation period  have significant consequences for the vertical extent of clouds. In Figure \ref{fig.tracer2zonalav} we show the time- and zonal-mean cloud mixing ratio as a function of pressure and latitude for models with four rotation periods.  With a given rotation period, the vertical extent of clouds is largest at the equator and decreases poleward.  The cloud thicknesses at the equator are almost the same among models with different rotation periods. At the poles, the zonal-mean  cloud thickness is smaller when the rotation period is shorter.  For rapidly rotating models,  $f$ spans a larger range and is responsible for the greater change of the zonal-mean cloud thickness. 
As a result of the equator-to-pole cloud thickness variation, the time- and zonal-mean temperature-pressure (T-P) structures exhibit a systematic equator-to-pole variation  due to the cloud radiative feedback.  Figure \ref{fig.zonalmeantp} shows the time-averaged zonal-mean T-P profiles at selected latitudes for the model with a rotation period of 2.5 hours. The isobaric temperature variation can reach more than 100 K from the equator to the poles.  On top of this systematic equator-to-pole zonal-mean variation,  instantaneous differences in the T-P structures associated with  the cloudy and clear-sky regions are also present, similar to that shown in \cite{tan2020bd}.
 
\subsubsection{Equatorial wave properties} 
Dynamics near the equator shows qualitative differences to that at mid-to-high latitudes, which can be seen from two features. The first  is the morphology of the eddies. Eddies near the equator are likely more elongated in the zonal direction,  which is in contrast to the horizontally isotropic vortices  at mid-to-high latitudes. This feature is more prominent in models with rotation periods of 5 and 2.5 hours.   The second difference is the time evolution of the eddies. Equatorial eddies exhibit systematic eastward or westward propagation, while
trajectories of mid-to-high-latitude vortices  are more akin to random walks and show no systematic migration along certain directions. Figure \ref{fig.equatortime} shows the Hovm$\ddot{\rm o}$ller diagrams (longitude-time sequence) of outgoing thermal flux at the equator as a function of longitude and time for models with different rotation periods. All models exhibit obvious characteristic eastward propagating patterns  (those moving towards the upper right of the panels) with zonal speeds on the order of a few hundred $\mps$. On the contrary, Figure \ref{fig.equatortime45} in Appendix \ref{ch.hovmoller} similarly shows the Hovm$\ddot{\rm o}$ller diagrams at $45^{\circ}$ latitude where stochastically evolving vortices dominate, and there is not evidence of eastward or westward propagation of the eddies.  These propagating eddies at low latitudes have much larger zonal wavelengths than the mid-to-high-latitude vortices (comparing Figures \ref{fig.equatortime} and \ref{fig.equatortime45}), showing dominant features characterized by zonal wavenumbers of a few. Westward propagation of eddies is also present with a slower phase speed.  The time evolution of the zonally propagating eddies is not always coherent, and often shows complications as they evolve. For example, some existing perturbations disappear when they propagate while some new perturbations are generated.  
The propagating eddies at low latitudes are likely equatorially trapped waves, and they  exert significant effects on the short-term evolution of lightcurve variability as will be discussed in Section \ref{ch.lightcurve}. 

Based on both the morphology and time evolution of the eddies, latitudinal boundaries that separate the two groups of atmospheric motions seem to emerge. The boundary is closer to the equator with shorter rotation period.  This boundary is likely related to the equatorial  deformation radius $L_{\rm eq}= \sqrt{c_g/\beta}$. Poleward of this distance from the equator, there is sufficient room for mature vortices to develop and the dynamics is dominated by vortices. Within this scale from the equator,  dynamics are shaped by  equatorially trapped waves that are coupled to the cloud radiative effect. {\btt Using $c_g\sim 2000\mps$ obtained from fitting theoretical off-equatorial deformation radius to the measured sizes of dominant vortices as a function of $f$ in the constant$-f$ models of \cite{tan2020bd}, the equatorial deformation radius $L_{\rm eq}$ extends to about $24^{\circ}$, $17^{\circ}$, $12^{\circ}$ and $9^{\circ}$ away from the equator for models with rotation periods of 20, 10, 5 and 2.5 hours, respectively.} This is roughly consistent with that shown in Figure \ref{fig.global_tq} and \ref{fig.global_flux} and their time evolution (not shown).

\begin{figure*}
	\includegraphics[width=2\columnwidth]{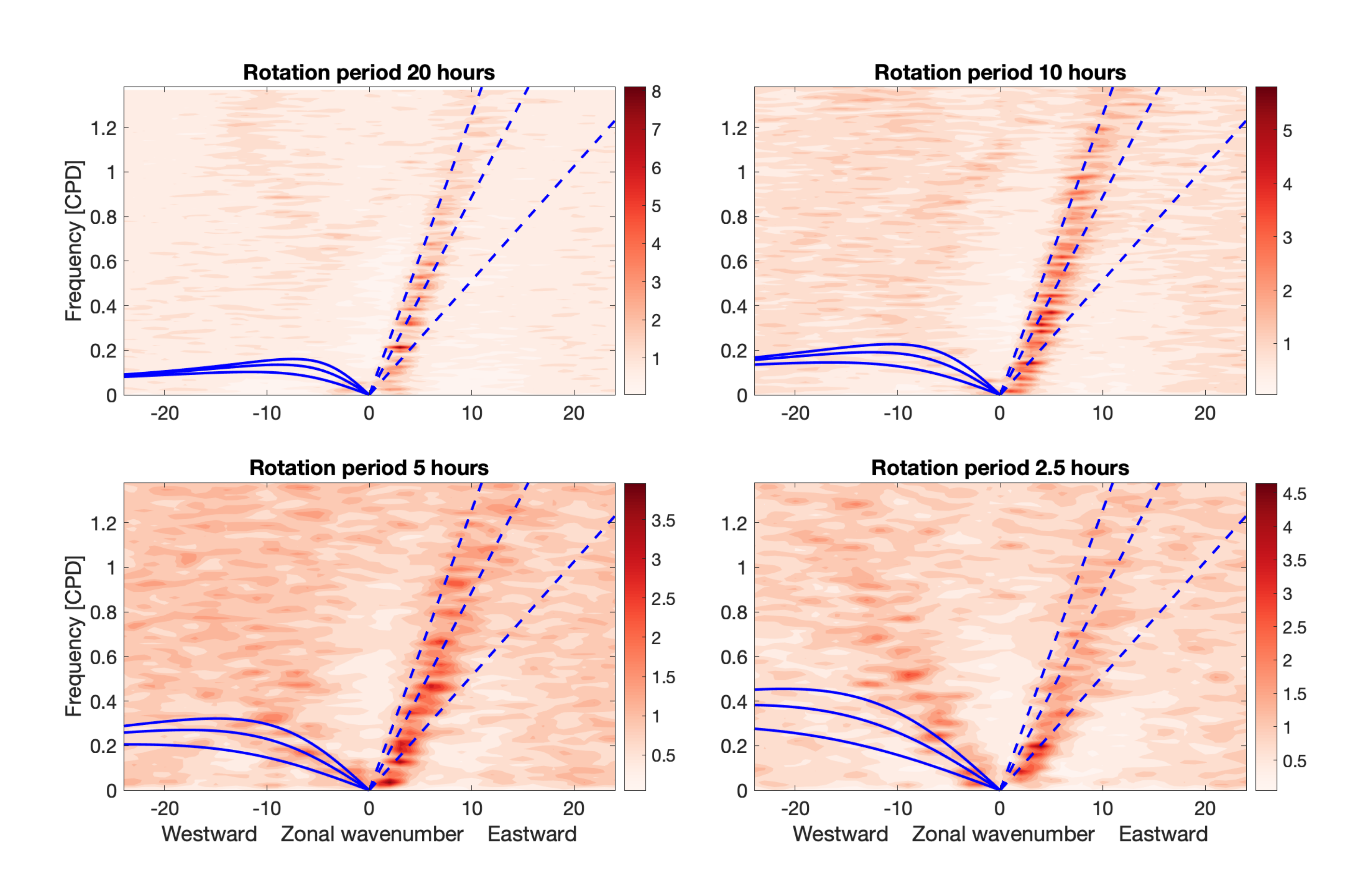}
    \caption{Wavenumber-frequency relative power spectrum of the symmetric components around the equator for models with four rotation periods of 20, 10, 5 and 2.5 hours and with a drag timescale $\tdrag=10^5$ s. Solid lines are three sets of $n=1$ equatorially trapped Rossby waves with three equivalent depths of $h_e=300$, 150 and 50 m (equivalent to  Kelvin wave phase speeds of 548, 387 and 224$\mps$). {\btt Here $n$ is the meridional mode number used in classic equatorial wave theory.} Dashed lines are three sets of Kelvin waves with the same equilibrium depths. The horizontal axis represents zonal wavenumber with negative values representing westward propagation and positive values representing eastward propagation. The vertical axis is frequency in a unit of cycle per day (CPD). {\btt The relative power spectrum shown here is non dimensional. Fluxes used in this analysis are sampled every 0.3 day.} }
\label{fig.spacetime_symm}
\end{figure*}

\begin{figure*}
	\includegraphics[width=2\columnwidth]{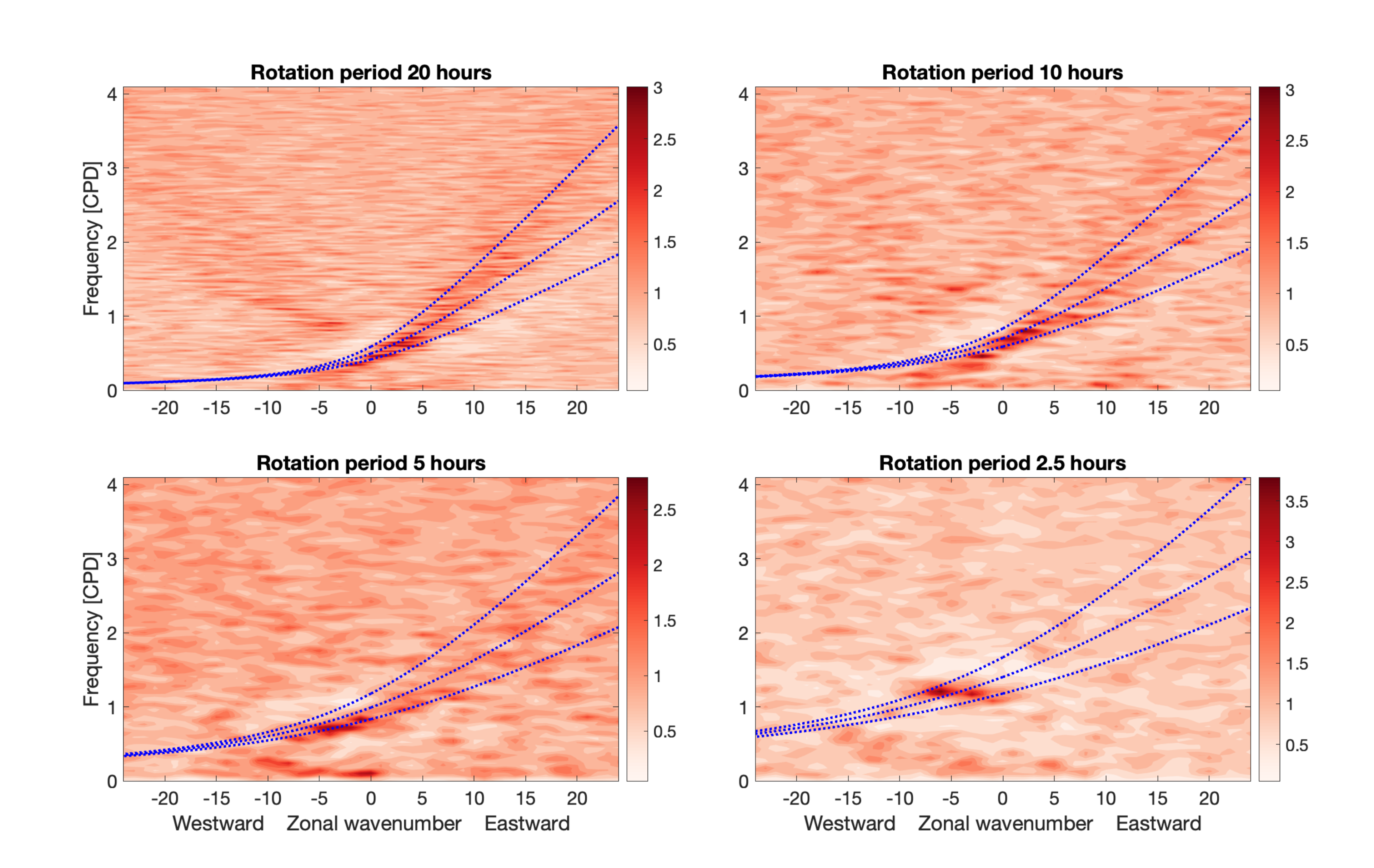}
    \caption{Wavenumber-frequency relative power spectrum of the anti-symmetric components around the equator for the same data sets used in Figure \ref{fig.spacetime_symm}. Dotted lines are three sets of MRG waves with three equivalent depths of $h_e=400$, 200 and 100 m (equivalent to having Kelvin wave phase speeds of 632, 447 and 316$\mps$). {\btt Fluxes used in this analysis are sampled every 0.1 day.}
    }
\label{fig.spacetime_antisym}
\end{figure*}

Now we turn to characterize properties of equatorial waves that are present in our simulations. We perform a spectral analysis at the equator in the wavenumber-frequency domain, similar to that performed in \cite{wheeler1999} and \cite{showman2019}.  The brief procedure is the following.  We  perform two-dimensional fast Fourier transforms on the outgoing thermal flux as a function of longitude and time near the equator to obtain the raw Fourier coefficients in the wavenumber-frequency space. These raw coefficients are then heavily smoothed in the wavenumber-frequency space to generate the background spectrum. Finally,  the raw coefficients are divided by the background spectrum to obtain the relative power spectrum, in which significant signals will appear to have values greater than 1. The eddy fields are further decomposed  into symmetric and anti-symmetric components about the equator because it helps to clarify the wave properties \citep{wheeler1999,kiladis2009}. The relative power spectrum  of models with different rotation periods are  shown in Figure \ref{fig.spacetime_symm} for the symmetric components and Figure \ref{fig.spacetime_antisym} for the anti-symmetric components.

Our space-time spectral analysis demonstrates the robust existence of groups of zonally propagating waves. In the symmetric components, a group of eastward waves with  zonal wavenumbers between 2 to 10 are present in all models. Weaker signals of westward waves with zonal wavenumbers between $-10$ and $-2$ exist in models with rotation periods of 5 and 2.5 hours but not in those with rotation periods of 20 and 10 hours. In the anti-symmetric components, models with rotation periods of 20 and 10 hours exhibit evidence of both eastward and westward waves with zonal wavenumbers in between about $-5$ to $5$, and those with 5 and 2.5 hours host westward waves with zonal wavenumbers between $-10$ to $1$. 

We compare the signals in Figure \ref{fig.spacetime_symm} and \ref{fig.spacetime_antisym} to dispersion relations of adiabatic, freely propagating equatorial waves derived from the shallow-water system \citep{matsuno1966}. These  waves are characterized by an equivalent depth $h_e$, such that the gravity wave speed (which is the same as the Kelvin wave speed) $c_g=\sqrt{gh_e}$ where $g$ is the surface gravity.\footnote{Linear equatorial wave theories  in the continuously stratified atmospheres  often decompose the atmosphere into a set of shallow-water systems with different equilibrium depths  corresponding to different vertical modes of the continuously stratified atmosphere (see  discussion in, e.g., \citealp{kiladis2009,tsai2014}).  This is partly why shallow-water equatorial waves are often used as a baseline for comparisons to either observations or simulated atmospheres (see a review by \citealp{kiladis2009}). } Although waves in our simulations are highly diabatic and  tightly coupled to radiative effects of clouds, this serves as a baseline comparison. Solid lines  in Figure \ref{fig.spacetime_symm} represent the dispersion relations of adiabatic free equatorial Rossby waves with a meridional mode number $n=1$ and with three equivalent depths of 300, 150 and 50 m (equivalent to gravity wave phase speeds of 548, 387 and 224$\mps$). Dashed lines in Figure \ref{fig.spacetime_symm} represent dispersion relations of Kelvin waves with the same equivalent depths. In the anti-symmetric components  shown in Figure \ref{fig.spacetime_antisym}, $n=0$ westward mixed Rossby-gravity (MRG) waves and $n=0$ and eastward inertia gravity (EIG) waves  with three equivalent depths of 400, 200 and 100 m (gravity wave phase speed of 632, 447 and 316$\mps$) are plotted as dotted lines.

In the symmetric components (Figure \ref{fig.spacetime_symm}),  evidence of  Kelvin waves are quite strong for all models as the spectral powers follow the Kelvin wave dispersion relations (dashed lines) reasonably well. Amplitudes of the Kelvin waves  are stronger among low zonal wavenumbers between $2-6$. The wave speeds of our simulated Kelvin waves are somewhat dispersive. Kelvin modes with lower zonal wavenumbers (longer zonal wavelengths)  typically have slower phase speeds than those with higher zonal wavenumbers. A phase speed of $\sim550\mps$ brackets the upper limit of phase speeds in all models. The phase speeds of waves with a zonal wavenumber of 2 in cases with rotation period of 10 and 5 hours are  much slower than $220\mps$. There is no evidence of equatorial  Rossby waves in models with rotation periods of 20 and 10 hours, and the evidence is tentative in models with 5 and 2.5 hours. However, even in the case with 2.5 hours, the westward branch  with relatively high frequencies $>0.5$ cycle per day (CPD) obviously deviates from the theoretical Rossby-wave dispersion relation.
There are no wave signals at frequencies higher than those shown in Figure \ref{fig.spacetime_symm} in the symmetric components, suggesting the absence of high-frequency inertia gravity waves.  

In the anti-symmetric components (Figure \ref{fig.spacetime_antisym}),  MRG and $n=0$ EIG waves are  evident in cases with rotation periods of 20, 10 and 5 hours. They have either eastward or westward phase velocities with  speeds larger than that of the Kelvin waves.  In the case of  2.5 hours, although the wave signals lie in between the dispersion relations of theoretical MRG waves assuming different equilibrium depth, the  signals show somewhat constant frequency between zonal wavenumber -7 to -3, making the  interpretation less obvious. Nevertheless, these diagrams (Figure \ref{fig.spacetime_symm} and \ref{fig.spacetime_antisym}) quantify the propagation of waves and help to recognize the wave types.

\begin{figure}
	\includegraphics[width=1\columnwidth]{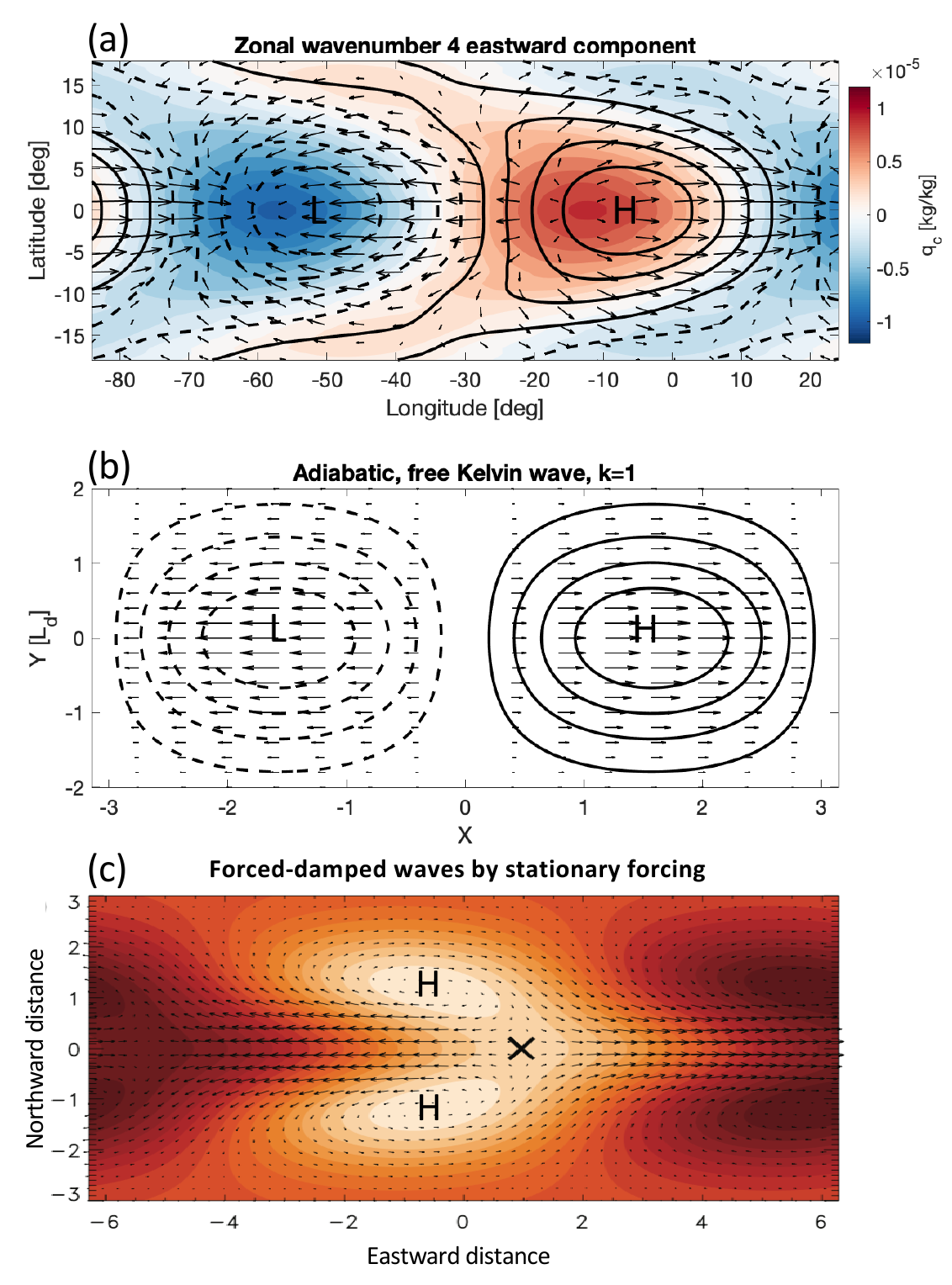}
    \caption{
    {\bf Panel  (a)}: a snapshot of the horizontal structure of eastwardly propagating, symmetric wave at 0.23 bar with a zonal wavenumber 4 and a frequency  of about 0.15 CPD in the simulation with a rotation period of 10 hours. Thick solid lines are positive geopotential anomalies from $0.5\times10^4$ to $3.5\times10^4\;{\rm m^2s^{-2}}$ and thick dashed lines are negative geopotential anomalies from $-3.5\times10^4$ to $-0.5\times10^4\;{\rm m^2s^{-2}}$ (with H marking the high-geopotential center and L marking low-geopotential center). Arrows represent wind vectors, and colours represent cloud mixing ratio associated with this wave. {\bf Panel (b)}: an example of an adiabatic free Kelvin wave as a function of non-dimensional zonal and meridional distances. Similar to panel (a), arrows represent wind vectors and thick lines represent height perturbations in the shallow-water system. {\bf Panel (c)}: An analytic forced-damped stationary wave solution forced by a zonal-wavenumber-1 stationary forcing centered around (0,0). Light colours represents larger values in the height field and arrows are eddy wind vectors.  This is adopted from Showman \& Polvani (2011),  which the details of the solution is referred to.
    }
\label{fig.kelvinwave}
\end{figure}

{\btt Because eastward Kelvin waves seem dominant in both the equatorial time evolution of out-going thermal flux (Figure \ref{fig.equatortime})  and  the spectral analysis (Figure \ref{fig.spacetime_symm}),} we further illustrate their properties in the physical domain by extracting from the total dataset through filtering in the wavenumber-frequency domain, similar to the procedure in \cite{wheeler1999}. At a given latitude, we choose a signal with a zonal wavenumber and a frequency  in the wavenumber-frequency domain, then project it back to the longitude-time domain. We repeat this process for all latitudes near the equator and obtain the propagating wave in the physical space.  Panel (a) in Figure \ref{fig.kelvinwave} shows the snapshot of an eastward symmetric wave with a zonal wavenumber of 4 and a frequency of about 0.15 CPD that shows a strong signal in the model with a rotation period of 10 hours (the upper right panel in Figure \ref{fig.spacetime_symm}). An example of adiabatic free Kelvin wave is shown in panel (b) as a comparison to our simulated wave. The adiabatic free Kelvin wave has eddy zonal velocity in phase with the pressure anomaly, and the disturbances decay away from the equator. Our simulated wave shows certain similarities to the adiabatic free Kelvin wave --- velocities have much larger zonal components than the meridional components in most places, and the geopotential anomalies decay away from the equator.  Note that cloud abundances correlate well with the convergence/divergence of the wind field (see a comparison between vectors and the colors that represent the cloud abundance). 

Meanwhile, in the simulated wave, the eastward zonal velocity pattern  exhibits a moderate eastward phase shift compared to the geopotential anomalies in the zonal direction (whereas those in the adiabatic free Kelvin wave are in-phase). Moreover, the  simulated wave exhibits a northwest-southeast phase tilt in the northern hemisphere and a southwest-northeast phase tilt in the southern hemisphere for the geopotential anomalies. This likely reflects a tendency that the circulation  resembles wave patterns excited by a stationary heat source that is symmetric about the equator \citep{matsuno1966,gill1980}. An example of such a stationary wave solution subjected to damping on both velocity and layer thickness in the shallow-water system adopted from \cite{showman2011} is shown in panel (c). This so-called Matsuno-Gill pattern is characterized by an eastward shift of the Kelvin component at the equator and a westward shift of Rossby components off the equator. Therefore it shows the northwest-southeast phase tilt in the northern hemisphere and a southwest-northeast phase tilt in the southern hemisphere, as well as that the maximum positive zonal velocity is east of the height maximum at the equator. In our simulations, when cloud forms by convergence at the equator, the accompanied heat source drives the flow towards the Matsuno-Gill shape. However, because clouds  themselves are also advected by flows over a timescale that is comparable to that required to form the Matsuno-Gill pattern, the advection of the heat source disrupts the complete formation of the Matsuno-Gill pattern. This is why our simulated wave still shows quantitative discrepancies to that in panel (c).

It is likely that our simulated eastward, Kelvin-like waves resemble properties of both the adiabatic free Kelvin waves and the forced-damped Matsuno-Gill circulation. Recently, using a quasi-linear moist shallow-water system, \cite{vallis2020} showed that the Matsuno-Gill-like circulation  is excited by latent heating (which itself is  coupled to the flow, somewhat analogous to the cloud radiative heating in our cases) at the equator, and  a slow eastward migration of the whole pattern is driven by the interactions of the flow and moisture field. It is unclear how much the mechanism shown in \cite{vallis2020} participates in the eastward propagation of our simulated waves in addition to that related to the free Kelvin wave. Teasing out the detailed mechanisms in our simulations is {\btt beyond the scope of this paper} as they are highly nonlinear. Here we still refer  these zonally traveling disturbances as waves, partly because these disturbances follow wave dispersion relations reasonably well (as shown in Figures \ref{fig.spacetime_symm} and \ref{fig.spacetime_antisym}).

The origin of the equatorial waves in our simulations is likely related to self-excitation by cloud radiative feedbacks. However, even treating them as waves, mechanisms controlling the wave properties are unclear and little previous work exists.   Linear stability analyses on the cloud radiatively coupled dynamics was carried out for inertia gravity waves in a $f$-plane system and a quasi-geostrophic system by \cite{gierasch1973}, and they showed that unstable modes are possible as a result of cloud radiative instability. Here, we extend their theory to the equatorial $\beta$-plane  which is appropriate for equatorially trapped waves. The derivations are shown in   Appendix \ref{ch.theory}. In the linear theory, we find that unstable modes are possible for a set of Kelvin and $n=0$ MRG modes, suggesting a source of kinetic energy on the equatorial waves shown in our models.  However, the unstable modes from the theory do not propagate. All other modes that resemble the classic adiabatic free equatorial waves discovered by \cite{matsuno1966}, including propagating Kelvin, Rossby, MRG, eastward and westward inertia gravity modes,  show damping on the eddy amplitudes due to thermal radiation.   The linear theory is not a total failure because it still predicts the kinetic energy sources for most equatorial waves in our simulations,  but it cannot explain the  propagation of the simulated waves. This situation is similar to that  in the non-rotating two-dimensional system shown in \cite{tan2020bd}. 

The wave speeds shown in Figure \ref{fig.spacetime_symm} and \ref{fig.spacetime_antisym} are  significantly slower than expected from adiabatic waves with long vertical wavelengths in conditions appropriate for our simulated atmospheres, the latter of which  is characterized by a much larger dry equivalent depth $h_{e, {\rm dry}}=c^2_{g,{\rm dry}}/g\sim 4000$ m, where $c_{g, {\rm dry}} \approx2NH\sim2000 \mps$, $N$ is the Brunt-Vaisala frequency and $H$ is the scale height. This is similar to equatorial waves in Earth's troposphere which are affected by latent heat released from moist convection \citep{wheeler1999}. To understand the reduced phase speed of tropical waves by diabatic effects, the following idealized framework has been proposed (see a review by \citealp{kiladis2009}). Suppose that the large-scale diabatic heating and cooling $Q$ is included in the linearized thermodynamics system as the following
\begin{equation}
    \frac{\partial}{\partial t}\left(\frac{\partial \phi}{\partial z}\right )+wN^2=Q,
    \label{eq.linear_therm1}
\end{equation}
where $\phi$ is the geopotential (note that $\partial\phi/\partial z$ is proportional to the temperature perturbation), $z=-H\log(p/p_s)$ is the log-pressure coordinate, $p$ is pressure, $p_s$ is a reference pressure, and $w$ is vertical velocity in log-pressure coordinates. Furthermore, if the heating and cooling are proportional to the vertical velocity such that $Q=\alpha w N^2$ (where $\alpha$ is an arbitrary constant), then Equation (\ref{eq.linear_therm1}) becomes
\begin{equation}
    \frac{\partial}{\partial t}\left(\frac{\partial \phi}{\partial z}\right )+w(1-\alpha)N^2=0.
\end{equation}
If $\alpha$ is positive and less than 1,  diabatic heating and cooling effectively reduce the stability of the atmosphere and therefore reduce the  wave phase speeds. For Earth's tropical waves, the problem reduces to determine  theoretical values based on parameterized moist convection for $\alpha$ to explain the observed wave speeds (e.g., \citealp{neelin1987,emanuel1994,haertel2004}), but the problem has not been completely solved \citep{kiladis2009}. For  waves coupled purely with cloud radiative feedback, the relation $Q=\alpha w N^2$ likely holds as well. In regions with upwelling, vapor is advected above the condensation level and clouds form, which generates warming near the condensation level due to cloud radiative effect. In regions with downwelling, cloud-free air is advected downward and clears out the region, enhancing the IR flux to space and inducing cooling. {\btt To quantitatively confirm the positive correlation between upwelling and heating, or downwelling and cooling, we calculate the cospectral power density for the quantity $w Q$ in the cloud forming layer, which is simply $2\mathbb{R}(w_k Q^{\ast}_k) $ where $w_k$ and $Q_k$ are the  coefficients at wavenumber $k$ space for  vertical velocity and heating rate, and $Q^{\ast}_k$ is the conjugate of $Q_k$. A detailed description of similar exercises in constant$-f$ models is referred to \cite{tan2020bd}.   The $w Q$ cospectral power density shows positive values for almost all zonal wavenumbers, proving that  $Q=\alpha w N^2$ with a positive $\alpha$  holds well in our models.} Therefore, we expect the above idealized framework may be used to qualitatively understand the reduced wave speeds seen in our simulations (however it does not address questions as why and how certain wave modes are excited but others are not). As for the detailed determination of $\alpha$, we defer to a future study.


\subsection{Results with weaker bottom drag}
\label{ch.drag}

\begin{figure}
	\includegraphics[width=1\columnwidth]{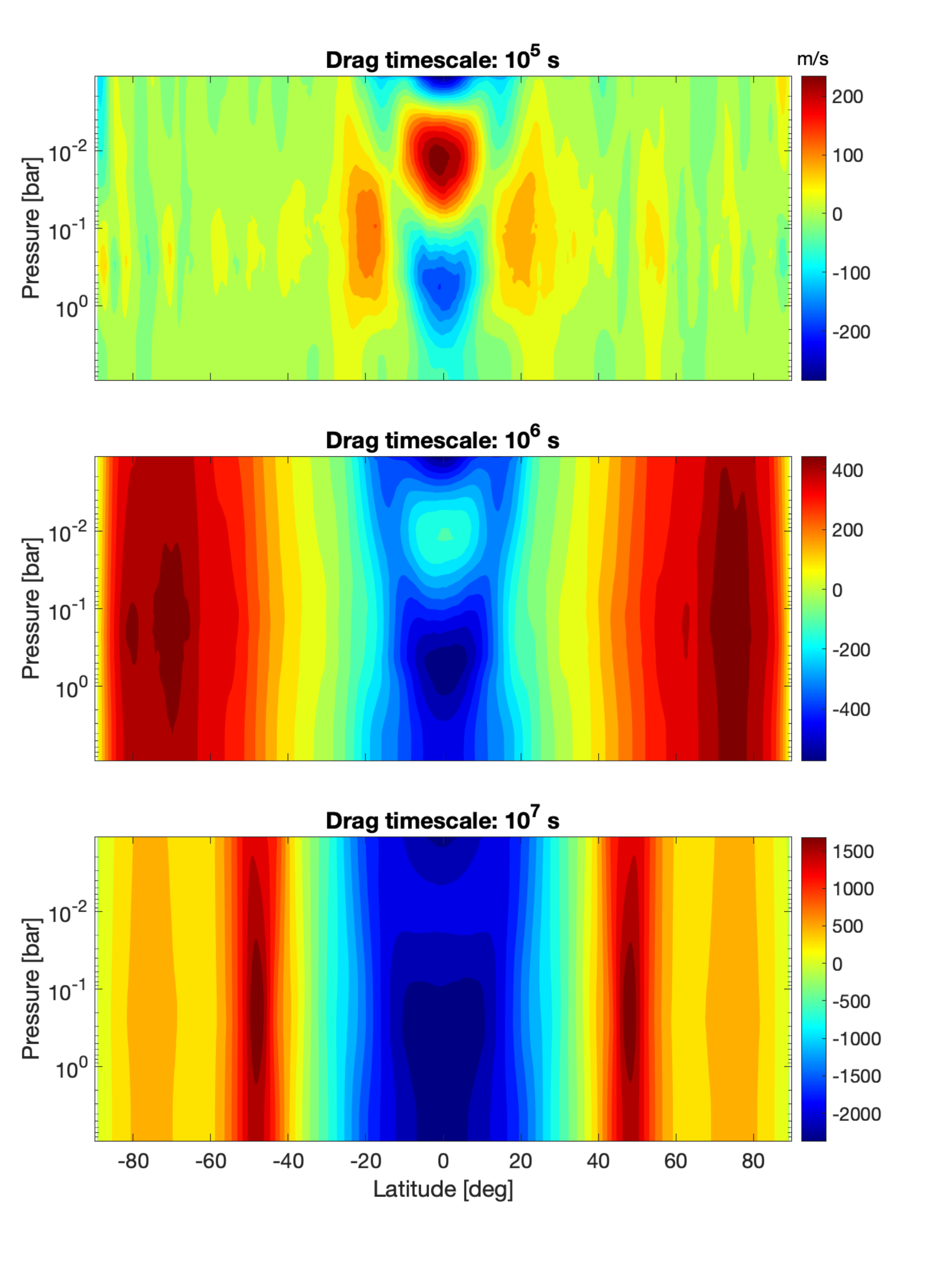}
    \caption{Time-averaged and zonal-mean zonal velocity as a function of pressure and latitude for models with a rotation period of 5 hours and  different  drag timescales $\tau_{\rm{drag}}=10^5$, $10^6$ and $10^7$ s from the top to bottom row.}
\label{fig.uzonalavr5}
\end{figure}

We examine properties of the circulation when the bottom frictional drag becomes weaker by performing two additional models with drag timescales of $\tdrag=10^6$ s and $10^7$ s and  a rotation period of 5 hours. Note that in all models, the frictional drag is applied only at pressures greater than 5 bars, well below the cloud condensation level of 0.5 bar. Figure \ref{fig.uzonalavr5} shows the time-averaged zonal-mean zonal wind as a function of pressure and latitude for models with  drag timescales of $\tau_{\rm{drag}}=10^5$, $10^6$ and $10^7$ s from the top to the bottom panel.  When the drag is strong ($\tdrag=10^5$ s), the zonal-mean zonal wind is much weaker than local horizontal wind speeds of vortices and turbulence. The zonal-mean zonal flow at low latitudes exhibits an interesting vertical wind shear, with a westward mean flow centering at around 0.5 bar and an eastward mean flow around 0.01 bar. {\btt The equatorial eastward jet centering around 0.01 bar corresponds to a superrotation, which requires up-gradient angular momentum transport to the equator by eddies. Our  analysis (not shown) suggests that horizontal eddy momentum transport by transient waves is responsible for this local superrotation. This  might be somewhat analog to those proposed for superrotation in  solar-system bodies, such as Venus, Titan, tropospheres of Jupiter and Saturn (see a recent review by \citealp{imamura2020}). }
In the model  with $\tdrag=10^6$ s, a broad equatorial westward jet and high-latitude eastward jets with speeds $\sim400\mps$ emerge. At low latitudes, the vertical  wind shear is similar to that with $\tdrag=10^5$ s despite the overall equatorial jet velocity being westward. The jet speeds are comparable to the horizontal mean eddy velocity, but the vortices are still nearly  isotropic at mid-to-high latitudes. Near the equator, the equatorial waves are Doppler shifted by the westward jet whose speed is comparable to that of wave propagation. In the model with $\tdrag=10^7$ s, the jet speeds become much stronger, reaching about $-2200\mps$ at the equator and more than $1500\mps$ at about $\pm50^{\circ}$ latitude. Compared to the jet structure with $\tdrag=10^6$ s, the meridional width of the equatorial jet with $\tdrag=10^7$ s is very similar and only the jet speed increases. However, at high latitudes, the cores of eastward jets are closer to the equator than those with $\tdrag=10^6$ s. There is a strong barotropic (pressure-independent) component for zonal jets with $\tdrag=10^7$ s. 

{\btt  The RMS horizontal eddy velocity in the cloud forming regions is about $400-550\mps$, which is much larger than  the  jet speeds in the model with $\tdrag=10^5$ s.  Even in the model with $\tdrag=10^6$ s, the jet speeds are only comparable to the eddy velocities.  In the following analysis, we focus primarily on the model with $\tdrag=10^7$ s in which the jet speeds well exceed the eddy velocities and the dynamics exhibits obvious differences to those with $\tdrag=10^5$ s. }

\begin{figure}
	\includegraphics[width=1\columnwidth]{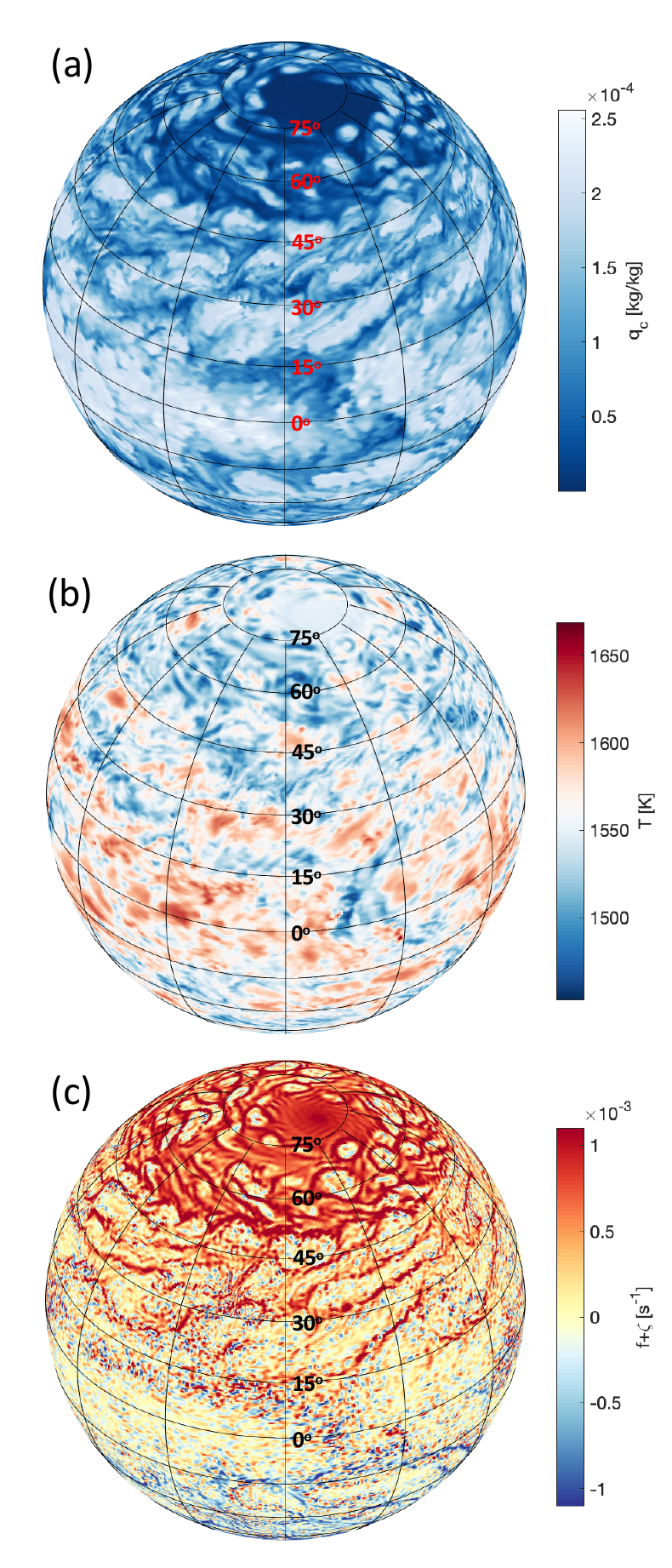}
    \caption{Instantaneous results from the model with a rotation period of 5 hours and a  drag timescale $\tau_{\rm{drag}}=10^7$ s after the model reached a statistical equilibrium state. Panel (a) shows the   cloud mixing ratio at 0.23 bar; panel (b) shows  temperature at 0.23 bar; and panel (c) shows the absolute vorticity at 0.23 bar. Note that strong eastward jets are located at $\pm48^{\circ}$ latitude. }
\label{fig.r5d7map}
\end{figure}

Cloud formation is  affected by the presence of strong zonal jets in the case with $\tdrag=10^7$ s.   Wind shears at the flanks of the mid-latitude jets create strong cyclonic regions poleward of the jets and  strong anti-cyclonic regions equatorward of the jets. Anti-cyclones which are associated with cloud formation tend to be more vulnerable in the strong cyclonic zones,  in the sense that they are   sheared apart and destroyed more easily than in other regions. On the other hand, formation of cyclones which are associated with thin clouds are more prevalent in the cyclonic zones. In addition, vortices tend to migrate towards regions with background absolute vorticity  closer to that of vortices (e.g., \citealp{scott2011,oneill2015}). There is a tendency that cyclones formed near the mid-latitude jets migrate to the poleward flanks of the jets where it is cyclonic, whereas anti-cyclones tend to migrate to the equatorward flanks of the jets. These behaviors are observed in time evolution of storms in the simulation with $\tdrag=10^7$ s (not shown). All these imply that, clouds associated with anti-cyclones are relatively depleted in the polar flanks of the mid-latitude jets and are more abundant in the equatorial flanks of the jets. {\btt The vortex behaviors influenced by the strong zonal jets described above may be one of the mechanisms responsible for  the cloud mixing ratio at 0.23 bar in panel (a) of Figure \ref{fig.r5d7map}. There are also zonal-mean upwelling equatorward of the mid-latitude jets where it is cloudy and zonal-mean downwelling poleward of the mid-latitude jets where it is less cloudy (not shown). This mean meridional circulation may or may not be a mechanism driving the meridional cloud gradient. It could simply be  radiatively driven {\it given} the existing meridional cloud gradient.  } Shapes of vortices are deformed due to the significant horizontal wind shear, especially near latitudes between $15^{\circ}-45^{\circ}$ and $50^{\circ}-70^{\circ}$ (see Figure \ref{fig.r5d7map}). Interestingly, there is a strong polar cyclone at each pole with the center of the cyclone not aligned with the pole. Clouds are also strongly depleted in the cyclones, and the same reasons mentioned above for cloud depletion poleward of the mid-latitude jets  might also be relevant. An instantaneous temperature map at 0.23 bar is shown in panel (b) and instantaneous absolute vorticity $f+\zeta$ is shown in panel (c) of Figure \ref{fig.r5d7map}, where $\zeta=\mathbf{k}\cdot \nabla_p\times\mathbf{v}$ is the relative vorticity, $\mathbf{v}$ is horizontal velocity vector, $\mathbf{k}$ is the local upward unit vector on the sphere and $\nabla_p$ is the horizontal gradient in pressure coordinates. Due to the radiative effects of clouds, strong cyclonic regions where clouds are relatively depleted are systematically colder than anti-cyclonic regions.

The development  of robust zonal jets with a weaker bottom frictional drag is consistent with that found in \cite{showman2019}, who studied atmospheric circulation of brown dwarfs using horizontally isotropic, randomly evolving thermal forcing. Storms and vortices excite Rossby waves, and their generation, propagation and wave breaking interact with the mean flow, driving zonal jets \citep{dritschel2008}. Without efficient removal of kinetic energy by the strong bottom frictional drag, zonal jets can be pumped up and maintained by wave-mean-flow interactions. Some features of the jets in our simulations are interesting. First, the jets are quite meridionally broad, with the equatorial westward jet extending to $\pm40^{\circ}$ latitude and subsequent eastward jets at about $\pm70^{\circ}$ latitude for the case with $\tdrag=10^6$ s and at about $\pm48^{\circ}$ latitude for the case with $\tdrag=10^7$ s. As a comparison, Jupiter  has about 7 subtropical zonal jets in each hemisphere (e.g., \citealp{ingersoll2004}).  Classic turbulence-driven jet theory predicts that the meridional jet spacing is related to wind speed via the Rhines scale $\pi\sqrt{U/\beta}$ where $U$ is a characteristic wind speed, indicating that the number of jet on a sphere is roughly given by $N_{\rm jet}\sim \sqrt{2\Omega a/U}$ where $\Omega$ is the rotation rate and $a$ is planetary radius (see reviews by, e.g., \citealp{vasavada2005, dritschel2008, showman2010}). Given a typical wind speed of several hundreds of $\mps$ in the case with $\tau_{\rm{drag}}=10^6$ s and $\sim 2000 \mps$ in the case with $\tau_{\rm{drag}}=10^7$ s, one would expect a number of zonal jets of $\sim9$ in the case with $\tau_{\rm{drag}}=10^6$ s and $\sim5$ in the case with $\tau_{\rm{drag}}=10^7$ s, which are obviously greater than that seen in our simulations. Second, the equatorial jets are westward in our simulations, whereas those in Jupiter and Saturn are eastward.

\begin{figure}
	\includegraphics[width=1\columnwidth]{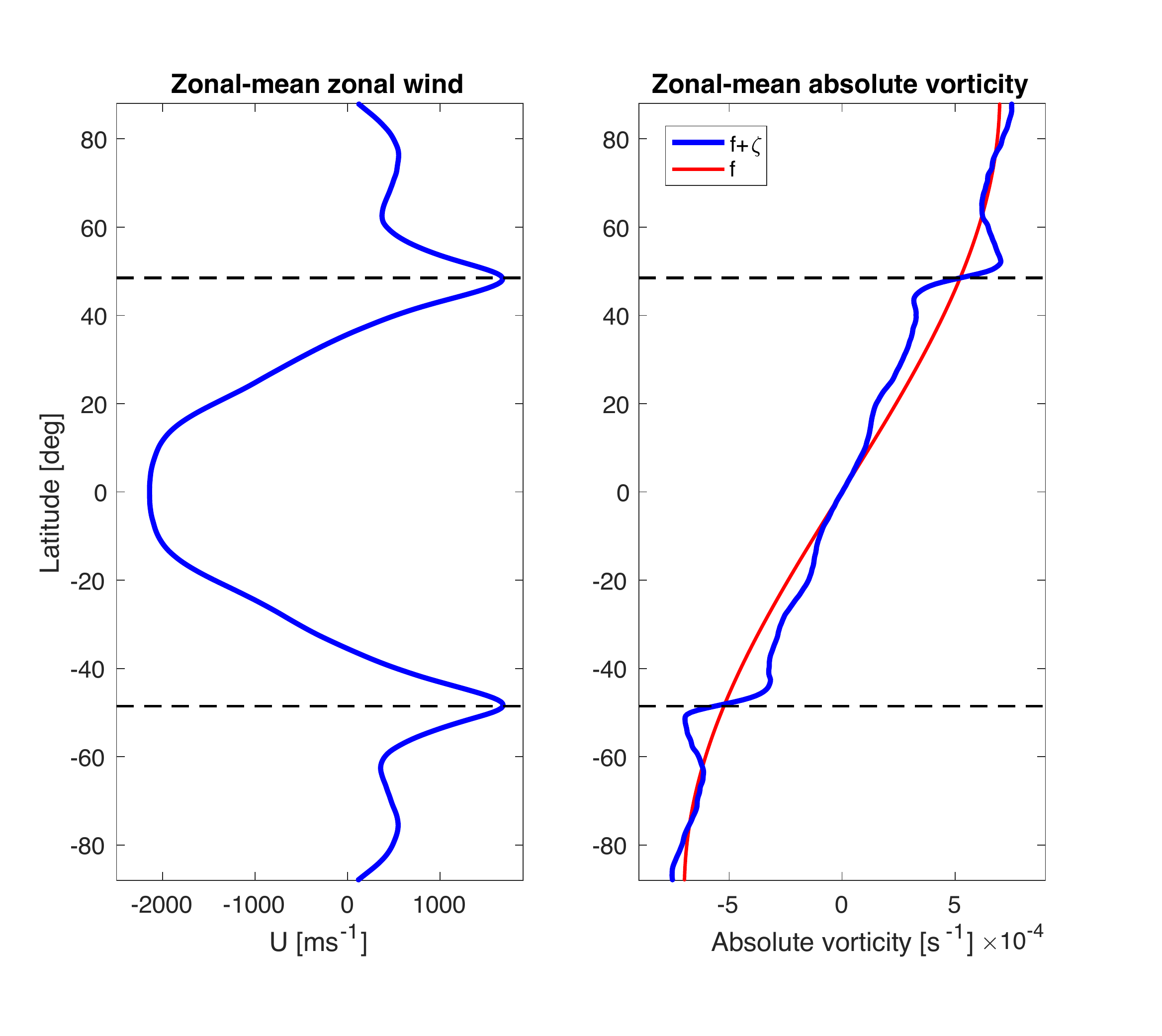}
    \caption{{\it Left panel:} time- and vertically averaged zonal-mean zonal wind as a function of  latitude for the model with a rotation period of 5 hours and  a drag timescale $\tau_{\rm{drag}}=10^7$ s after the model reached a statistical equilibrium state. {\it Right panel:} zonal-mean absolute vorticity ($f+\overline{\zeta}$) corresponding to the zonal-mean zonal wind in the left panel and planetary vorticity $f$.  }
\label{fig.ztmvorticity}
\end{figure}

The broader-than-expected jets are probably related to the efficient horizontal mixing of potential vorticity (PV) caused by strong vortices, which tends to break the PV staircase and smooth out the jet structure. One interpretation of the Rhines jet scaling is that the jets are a natural result of  PV staircases. Eastward jets correspond to the PV discontinuity whereas the westward jets correspond to the PV homogenization, and jet speed is determined by the meridional width of the PV staircase (e.g., \citealp{dritschel2008,scott2012}). But if the magnitude of vorticity sources is much larger than the background PV discontinuity, mixing of PV between the staircases leads to destruction of  the PV staircase, therefore the jet scaling does not apply well in this situation \citep{scott2012}.  In strongly forced and damped cases of \cite{showman2019}, the PV structure is disrupted by eddy vorticity and the resulting zonal jets are less sharp and meridionally broader than those that are  weakly forced   and damped. In our case, the vorticity mixing is even more extreme.   We find that in our simulations, the vorticity associated with mid-to-high-latitude vortices are typically much stronger than either the background planetary vorticity or that associated with the zonal jets. These vortices occurs somewhat randomly in space and time---their occurrence is not particularly constrained by the presence of zonal jets---and they can easily disrupt the zonal-mean PV structure. In panel (c) of Figure \ref{fig.r5d7map}, we can see that the eddy relative vorticity is far greater than the background planetary vorticity   in the case with $\tdrag=10^7$ s. Figure \ref{fig.ztmvorticity} shows the vertically averaged zonal-mean zonal wind in the left panel and its corresponding absolute vorticity (together with the planetary vorticity) in the right panel. Because the barotropic component dominates the jet structure in the case with $\tdrag=10^7$ s (panel (c) in Figure \ref{fig.uzonalavr5}), the absolute vorticity is a good representation of PV. The eastward jets correspond to the sharp gradient of absolute vorticity at $\pm48^{\circ}$ latitude. There is a lack of other absolute vorticity staircase in other regions. Meanwhile, absolute vorticity between $\pm40^{\circ}$ latitude tends to be homogenized (although still far from being completely well mixed). In some cases, the cross-equator homogenization of absolute vorticity can lead to a strong equatorial westward jet \citep{dunkerton2008}. Such a tendency may be responsible for the broad, strong equatorial westward jets in our simulations.

\begin{figure*}
    \includegraphics[width=2\columnwidth]{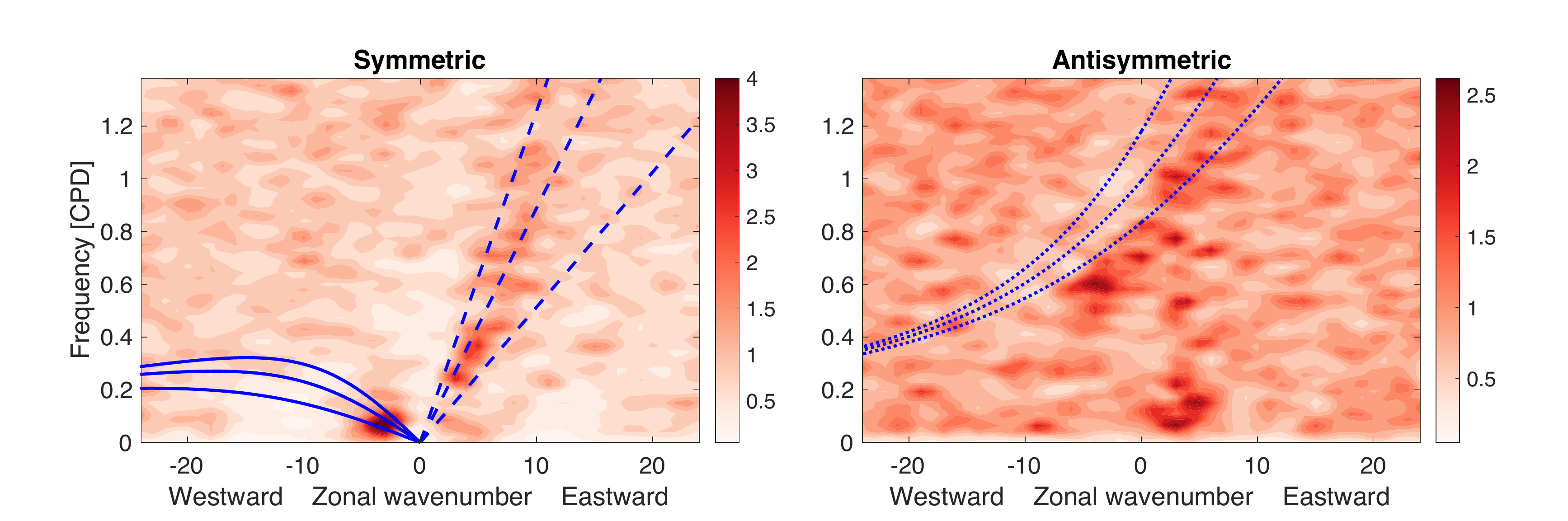}
    \caption{ {\btt Wavenumber-frequency relative power spectrum of the symmetric component around the equator (on the left panel) and of the anti-symmetric component (on the right panel) for the model with a rotation period of 5 hours and $\tdrag=10^7$ s. Doppler shifts in the power spectrum associated with a uniform zonal-mean zonal wind $-2130\;\mps$ are removed, and most signals shown in these panels are likely associated with waves propagating relative to the zonal-mean zonal wind. On the left panel, solid lines are theoretical equatorial Rossby waves and dashed lines are theoretical Kelvin waves, same as those shown in  Figure \ref{fig.spacetime_symm} for  a rotation period of 5 hours. On the right panel, dotted lines are theoretical MRG waves, same as those shown in Figure \ref{fig.spacetime_antisym} for a rotation period of 5 hours. } }
    \label{fig.spacetime_r5d7}
\end{figure*}

There is a local absolute vorticity maximum northward of the jet at $48^{\circ}$ latitude and a local minimum southward of the jet, indicating that the jet   violates the barotropic stability criteria. This is associated with the accumulation of cyclones northward of the jet and anti-cyclones southward of the jet. A positive feedback likely takes place in this case: the jet structure triggers accumulation of vortices, and the accumulation of those vortices further strengthens the jet structure that already promotes the accumulation. Similar influences of the migration and accumulation of vortices on the jet structure has been proposed in previous modeling studies of Jovian atmospheric dynamics \citep{thomson2016}. A balance may be reached  between the accumulation of vortices and the instability associated with the jet that tends to restore the jet structure towards stability. 

In box simulations with constant Coriolis parameter $f$ performed in \cite{tan2020bd}, we showed  that when the bottom frictional drag is weak ($\tau_{\rm drag}=10^7$ s), kinetic energy accumulates and a pair of a cyclone and a anticyclone forms with sizes comparable to the simulated domain size. In the global domain, the presence of $\beta$-effect  excites  Rossby waves, and eddy-mean-flow interactions channel the kinetic energy to the zonal direction, forming zonal jets instead of ever larger vortices. Our simulations demonstrate the importance of the $\beta$ effect on the formation of zonal jets. 

{\btt Finally, equatorially trapped waves also exist in the weak-drag models but their propagation is influenced by the equatorial jets. The influence is most prominent in the case with $\tdrag=10^7$ s, in which equatorial waves are Doppler shifted by the equatorial jet with a zonal-mean zonal wind $\sim  -2200\;\mps$. In the wavenumber-frequency analysis for this case,  we remove the Doppler shifts associated with the jet for which we assume a uniform zonal-mean zonal wind of $-2130\;\mps$ according to Figure \ref{fig.ztmvorticity}. The results are shown in Figure \ref{fig.spacetime_r5d7}. Part of the signals shown in these panels are likely associated with waves propagating relative to the zonal-mean zonal wind. We find evidence of eastward Kelvin waves and westward Rossby waves in the symmetric component around the equator. A family of MRG waves is also indicated in the analysis of the anti-symmetric component although their signals slightly deviate  from the theoretical ones.  However,  in reality, the equatorial jet has horizontal shears, whose influences on waves cannot be removed by simply assuming a uniform zonal-mean zonal wind in the analysis shown in Figure \ref{fig.spacetime_r5d7}. Such influences may be responsible for the abnormally strong  westward Rossby wave signal (which is only tentative in the strong-drag cases shown in Figure \ref{fig.spacetime_symm}) in the left panel, as well as  for signals around zonal wavenumber 3-5 and frequency 0.1-0.2 CPD in the right panel, which is absent in the strong drag cases (Figure \ref{fig.spacetime_antisym}). Indeed, when we slightly change the assumed uniform zonal-mean zonal wind in the wavenumber-frequency analysis, the position or strength of the above ``abnormal" signals is sensitive to the assumed zonal-mean zonal wind. But the signals associated with the Kelvin waves and MRG waves are only moderately affected, indicating that Kelvin waves and MRG waves seem robust.   }


\subsection{Simulated lightcurves}
\label{ch.lightcurve}

\begin{figure*}
	\includegraphics[width=1.8\columnwidth]{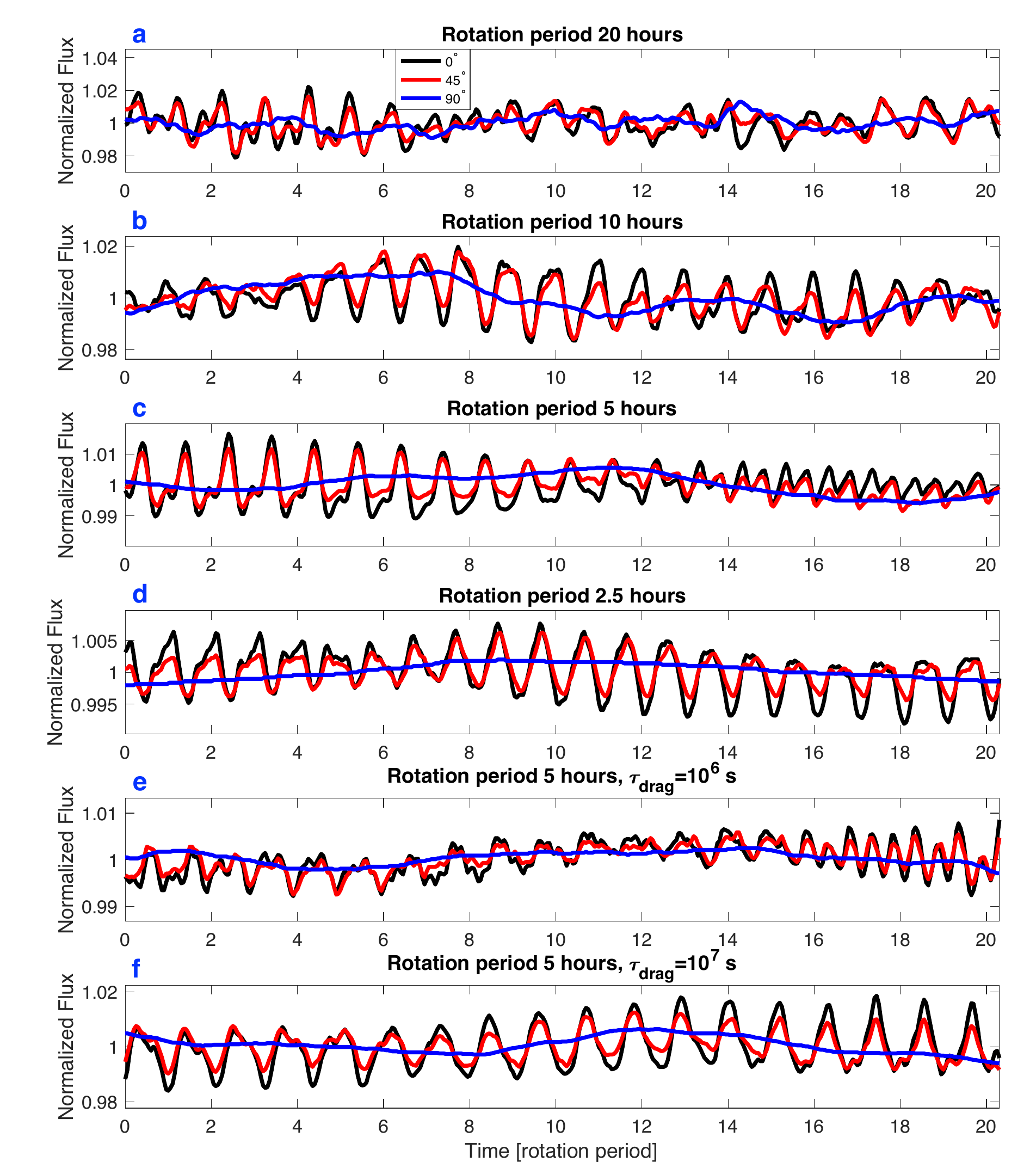}
    \caption{{\btt The top four panels (a to d) show normalized lightcurves  as a function of rotation period for models with four different rotation periods of 20, 10, 5 and 2.5 hours (from the top to bottom panels) and with a drag timescale $\tau_{\rm{drag}}=10^5$ s. In each panel, the black line represents a viewing angle of $0^{\circ}$ (equator-on), and the red line represents a viewing angle of $45^{\circ}$, and the blue line represents a viewing angle of $90^{\circ}$ (northern-pole-on). The bottom two panels show the normalized lightcurves for models with a rotation period of 5 hours and $\tdrag=10^6$ s in panel e and $\tdrag=10^7$ s in panel f (in which a strong westward equatorial jet develops).}}
\label{fig.lightcurve}
\end{figure*}


Due to the significant inhomogeneity in the outgoing thermal flux, lightcurve variability is expected from our simulations.  Figure \ref{fig.lightcurve} shows the  {\btt simulated lightcurves normalized to their time-mean values}  as a function of time that is normalized by the rotation periods of the models.  
Panels a to d are lightcurves from models with $\tdrag=10^5$ s and rotation periods of 20, 10, 5 and 2.5 hours, respectively. Panels e and f are lightcurves from models with a rotation period of 5 hours and $\tdrag=10^6$ and $10^7$ s, respectively. Black, red and blue lines in each panel represent  viewing angles of $0^{\circ}$, $45^{\circ}$ and $90^{\circ}$ relative to the equator, respectively---$0^{\circ}$ means  an equator-on view and $90^{\circ}$ means a northern-pole-on view. The former maximizes the rotational modulation of the lightcurve whereas in the latter there are no rotational effects. The atmospheres in our models are statistically symmetric between the northern and southern hemisphere, and therefore simulated lightcurves viewed from two hemispheres are qualitatively similar.

\begin{figure}
	\includegraphics[width=1\columnwidth]{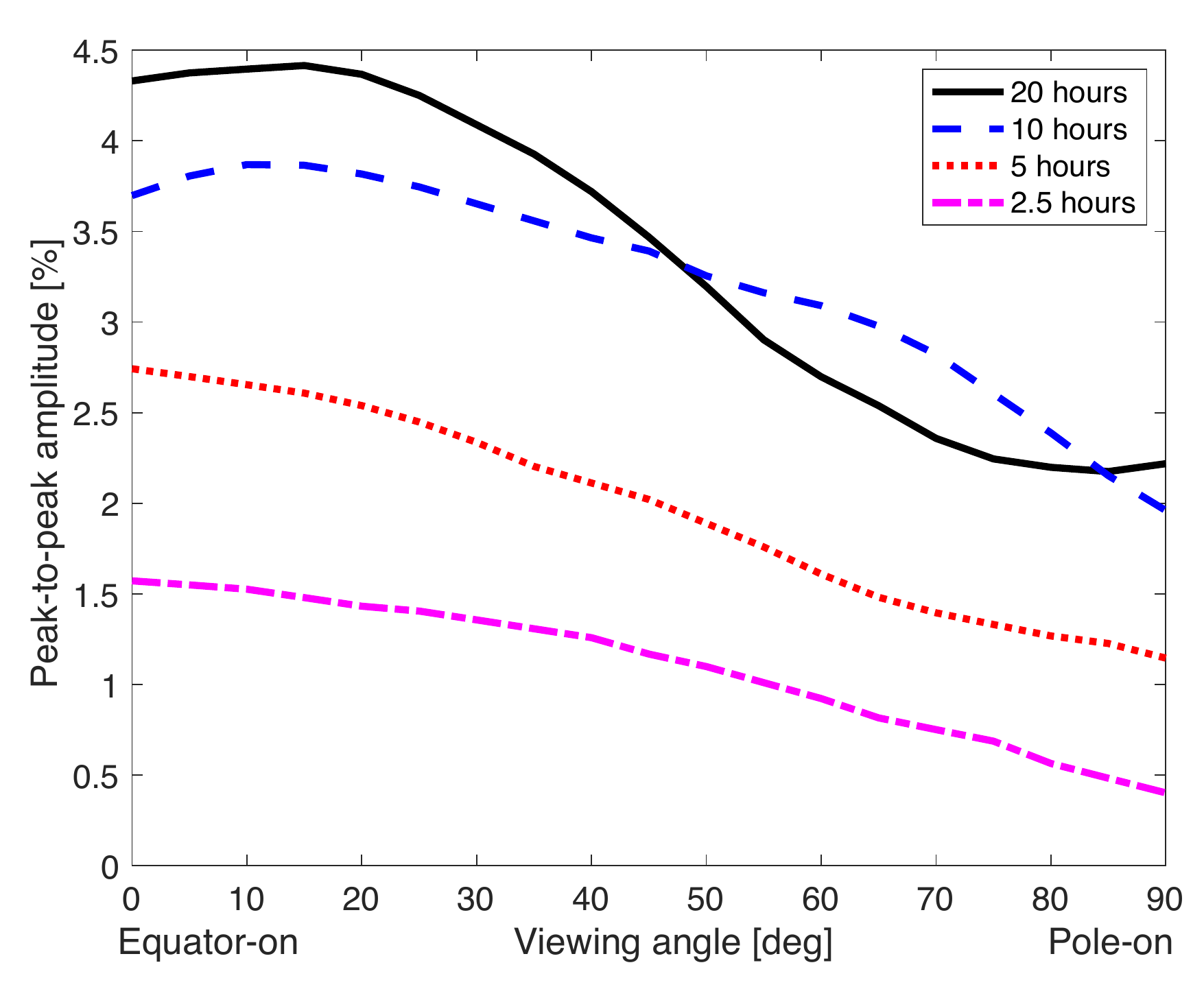}
    \caption{Peak-to-peak variability amplitudes for normalized lightcurves  with different viewing angles from models with different rotation periods and $\tdrag=10^5$ s. A subset of lightcurves are shown in Figure \ref{fig.lightcurve}. }
\label{fig.amplitude}
\end{figure}

The simulated lightcurves exhibit several important characteristics. For all models, the amplitudes of lightcurve variability are maximized when viewed equator-on, whereas they are  minimized when  viewed pole-on. Part of the reason is because the horizontal length scales of storms are larger at low latitudes, which causes larger flux perturbations when the object is viewed equator-on. This is consistent with the viewing angle dependence of near-IR flux variability amplitude found by \cite{vos2017}. The peak-to-peak amplitude of the equator-on lightcurve is typically a few percent and decreases with decreasing rotation period---the amplitude is almost 4\% for the case with rotation period of 20 and 10 hours, and slightly more than 2\% for the case with 5-hour rotation, and finally slightly more than 1\% for the case with 2.5-hour rotation. These characteristics are further summarized in Figure \ref{fig.amplitude}, which shows the peak-to-peak amplitudes of normalized lightcurves with different viewing angles.

Lightcurves with rotation periods of 10, 5 and 2.5 hours exhibit clear periodicity  related to the rotation period at least during certain times in their evolution. However, at some points the rotational periodicity can be complicated by the time evolution of the surface patchiness. For instance, in the case with a rotation period of 5 hours, the 5-hour periodicity is clear before  the 12th rotation period, but transition to double peaks within a rotation period after that. By eye, the shape of lightcurves for a rotation period of 20 hours is more complicated by short-term irregularities. This is perhaps  because the rotation period is long and the typical sizes of vortices are relatively large, which together indicate that evolution of individual vortices (typically over a timescale of several tens of hours) may significantly impact the lightcurve evolution over rotational timescales.

\begin{figure*}
	\includegraphics[width=1.8\columnwidth]{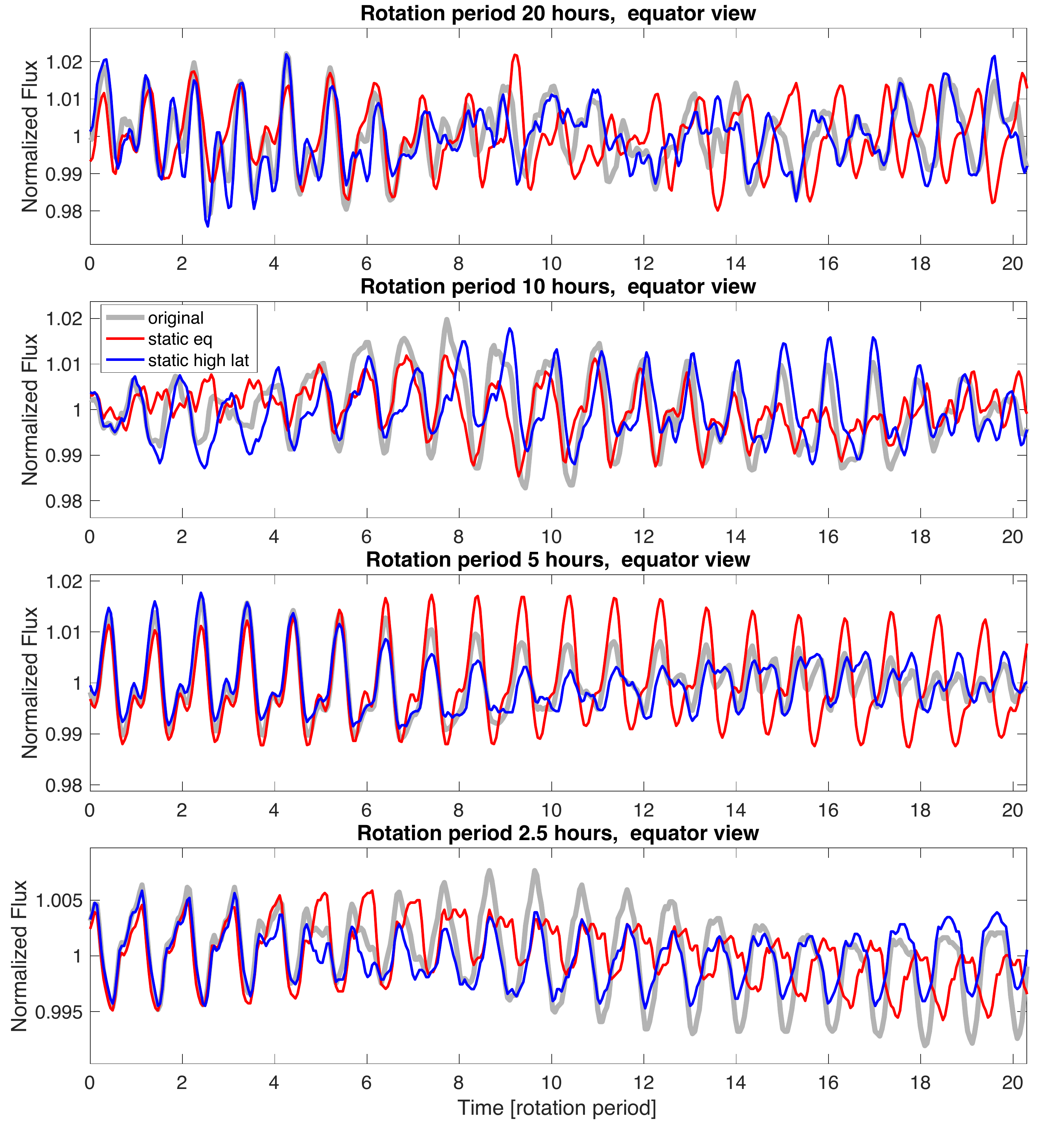}
    \caption{Normalized lightcurves with an equator-on viewing angle for models with four different rotation periods of 20, 10, 5 and 2.5 hours (from the top to bottom panels) and with a drag timescale $\tau_{\rm{drag}}=10^5$ s. In each panel, the thick grey line is the full-model lightcurve. The red line is from thermal flux outputs wherein the equatorial region is held fixed at time zero and stays invariant, while the mid-to-high-latitude regions are not modified. On the contrary, the blue line is from   outputs wherein the mid-to-high-latitude regions are held fixed at time zero. The equatorial regions here are defined as between $\pm 25^{\circ}$ for the model with a rotation period of 20 hours, and between $\pm 20^{\circ}$ for 10 hours, and between $\pm 15^{\circ}$ for both 5 and 2.5 hours.}
\label{fig.lightcurve_diag}
\end{figure*}

Viewed from pole-on, lightcurves are much smoother and show  much longer  periods of variation. The amplitude of pole-on lightcurves decreases  with decreasing rotation period. This lightcurve variability is caused by the statistical fluctuations of the ensemble of storms as discussed in \cite{tan2020bd}. The fewer storms projected in the disk, the larger the effects that the evolution of an individual storm can have on the the lightcurve. Sizes of storms near the poles inversely decrease with decreasing rotation period, therefore the above trend found in pole-on lightcurves emerges. Even though smaller than those viewed equator-on, the peak-to-peak amplitudes of pole-on lightcurves can still reach almost 2\% for cases with 20 and 10 hours and 1\% for the case with 5 hours, which are detectable given sensitivities of current instruments (e.g., \citealp{wilson2014,metchev2015}). This {\btt may contribute to}  some  long-term variations in observed lightcurves that are not easily explained by rotation modulation (e.g., some samples can be seen in \citealp{metchev2015}). 

{\btt Weak-drag models with $\tdrag=10^6$ and $10^7$ s similarly exhibit significantly time-varying waves and vortices, and some properties of their  lightcurves are similar to those with $\tdrag=10^5$ s, including peak-to-peak normalized lightcurve amplitude of a few percent, certain irregular time variability, viewing-angle dependence of variability amplitude,  and complications on the periodicity due to evolution of cloud patchiness.  However, because weak-drag models develop meridionally broad, strong zonal jets, zonal advection of clouds by the jets can modify the periodicity of the lightcurves. As shown in Figure \ref{fig.uzonalavr5}, the equatorial jets are both westward with  speeds of a few hundred $\mps$ and about $2000\mps$ in models with $\tdrag=10^6$ and $10^7$ s, respectively. Although Kelvin waves have eastward phase speeds relative to the mean flow,  in the rotational frame they travel to the west due to the strong westward equatorial jet. Their lightcurves, especially viewed equator-on, are expected to show periodicity longer than the rotational period of the model. This is most prominent in the case with $\tdrag=10^7$ s in panel f of Figure \ref{fig.lightcurve}, wherein the lightcurve  shows only 18 periods over 20 rotational periods.   }

We divide  contributions to the lightcurve variability by surface inhomogeneities into two groups of dynamical features: the zonally propagating equatorial waves and mid-to-high-latitude vortices (the latter do not migrate along the zonal direction).  We diagnose the effects of the two groups in the lightcurve by the following  process. First, in the model outputs, we artificially hold the equatorial region to be static. A synthetic lightcurve is generated based on this configuration, and the time evolution of the shape of the variability is caused only by evolution of mid-to-high-latitude vortices. Then, we artificially hold the mid-to-high-latitude regions to be static but do not hold the equatorial region static. The resulting time evolution of the shape of the variability represents only effects of the propagating equatorial waves. Figure \ref{fig.lightcurve_diag} shows these experiments for four models with $\tdrag=10^5$ s, in which the thick grey lines are the original full lightcurves; the red lines are lightcurves wherein the equatorial regions are held fixed from time zero; and the blue lines are lightcurves wherein mid-to-high latitudes are held fixed from time zero.  All cases are viewed equator-on. The equatorial region  is defined as in between $\pm25^{\circ}$ latitude for the case with rotation period of 20 hours,  between $\pm20^{\circ}$ latitude with 10 hours, and between  $\pm15^{\circ}$ latitude with 5 and 2.5 hours. {\btt These latitudinal bands are chosen  to safely include the equatorial trapped waves. Slightly changes of these latitudes do not influence our results and conclusions. }

In general, both the equatorial waves and mid-to-high-latitude vortices contribute to the short-term evolution of the full lightcurve variability---removing the time evolution of either component results in significant changes in the lightcurves. However, equatorial waves impact the time evolution of lightcurve shapes in more critical ways. First, the wave beating effect that causes the change of variability amplitude with time is much weaker when the equatorial regions are held fixed. This is most obvious in the case with a rotation period of 5 hours, in which the red curve (for which equatorial region is held time invariant) has  almost a constant amplitude, and the splitting to double sub-peaks  shown in the original lightcurve does not occur.    Second, there are significant phase differences in the lightcurve variability between the full lightcurve and the lightcurve with equatorial regions fixed. This is obvious in all cases and we take  the case with 2.5-hour rotation as an example: the red curve starts to show phase differences relative to the thick grey curve at a time of about the 7th rotation period; towards the end the red and grey curves show an almost $180^{\circ}$ phase difference. The blue curve, for which mid-to-high-latitude regions are held time invariant, mostly only show  differences in the local peaks and troughs of the variability, but not in the long-term evolution of the amplitude and the phase of the variability. Our diagnostic analysis suggests that equatorial waves have major impacts on the lightcurve variability and the time evolution of the shape of the variability, and mid-to-high-latitude vortices contribute to the local features of the lightcurves. This is in good agreement with the analysis of long-term observed lightcurves of a few BDs by \cite{apai2017}, in which they found that waves can explain the major evolution of lightcurves, and  ``spots'' are needed to fit the remaining local inconsistency between data and the wave model.

\begin{figure}
	\includegraphics[width=1\columnwidth]{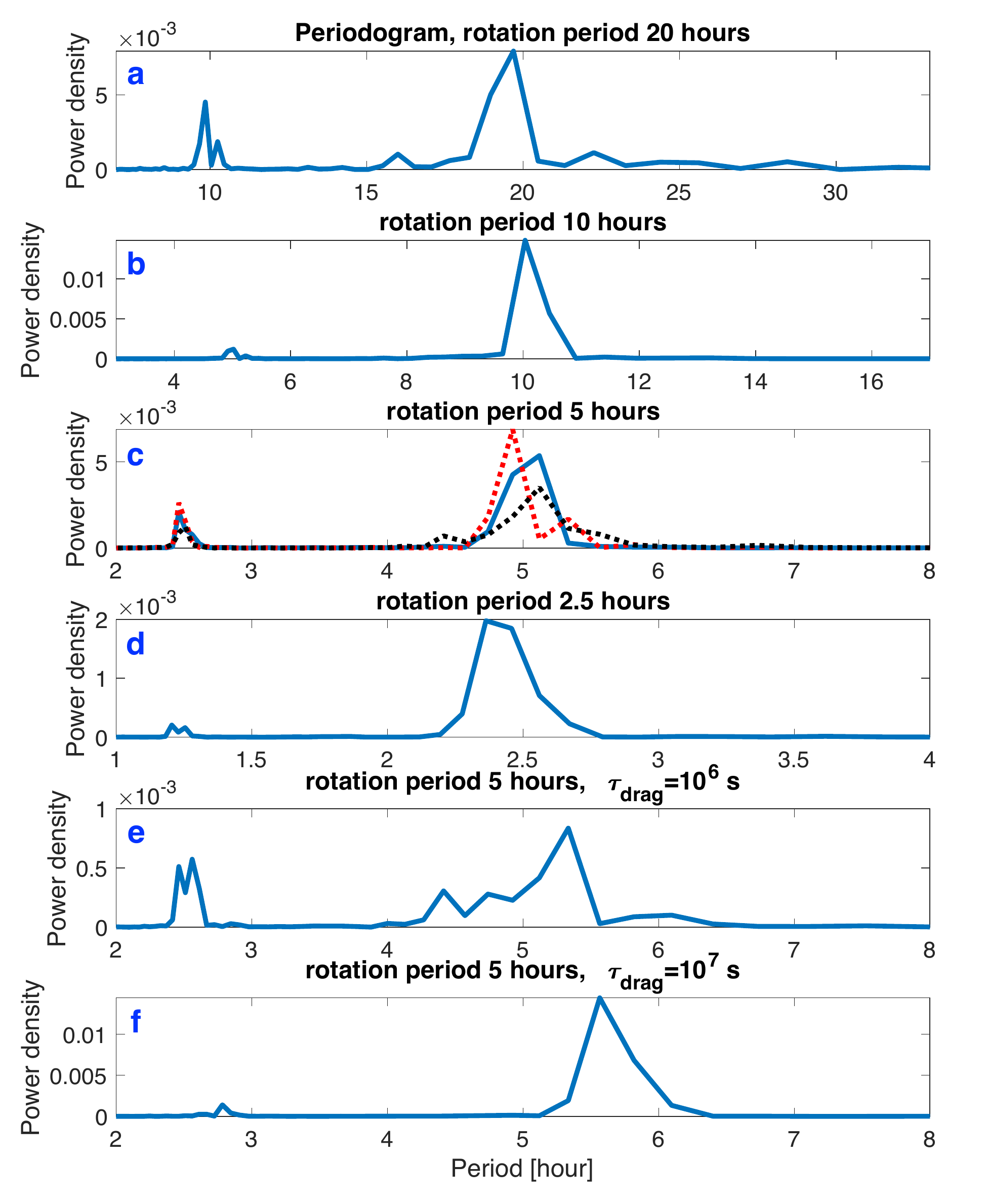}
    \caption{Periodograms of normalized lightcurves  for models with four rotation periods of 20, 10, 5 and 2.5 hours and with a drag timescale $\tdrag=10^5$ s (panel a to d). Periodogram for the model with a rotation period of 5 hours and $\tdrag=10^6$ s is shown in panel e, and for that with $\tdrag=10^7$ s is in panel f. In panel c, additional dotted lines are from output samples at different times much later than that for the solid blue line. }
\label{fig.periodogram}
\end{figure}

Both eastward and westward waves are present at low latitudes, and the eastward Kelvin waves are sometimes more dominant as visually shown in Figure \ref{fig.equatortime} and in the wavenumber-frequency analysis (Figure \ref{fig.spacetime_symm} and \ref{fig.spacetime_antisym}). This implies a faster rotation  of the equatorial features than the intrinsic planetary rotation, which may lead to a slightly shorter  period of the lightcurve variability than the underlying rotation period. This effect has been clearly shown in the case with 2.5-hour rotation in Figure \ref{fig.lightcurve_diag}, and now we quantify this using periodogram analyses of the simulated lightcurves shown in Figure \ref{fig.lightcurve}.  Figure \ref{fig.periodogram} shows power densities (in arbitrary unit) by the periodogram analysis for normalized lightcurves viewed equator-on for our models. 

Several features are interesting in the periodograms for models with $\tdrag=10^5$ s. First, some cases exhibit slight shifts of the major peaks  to a period slightly shorter than the rotation period.  In particular, cases with 20 and 2.5 hours show obvious shifts.   This is because of the eastwardly propagating Kelvin waves with  phase speeds of a few hundred $\mps$. 
Second, the power densities exhibit major peaks very close to (in some cases exactly the same as) the underlying rotation period of the models. However, these peaks are broadened, indicating that the dynamical features, either eastward or westward with various phase speeds, could  contaminate the periodicity of the lightcurve variability. Third, there is usually a secondary peak of the power density at a period of approximately half of the rotation period. This may be caused by the equatorial features with a zonal wavenumber 2 that appears twice to the ``observer" as the objects rotate once. {\btt
Finally, our models have shown that different equatorial waves that travel in different zonal velocities may contribute to the flux patchiness differently at different times (Figure \ref{fig.equatortime}). This could cause time variability on the properties of lightcurves when the lightcurves are sampled over a finite period (as the real observations do).   This is illustrated for the case with 5-hour rotation and $\tdrag=10^5$ s in panel c of Figure \ref{fig.periodogram}. We show two additional periodograms of  lightcurves that sample the model outputs at different times, and each of them also samples about 20 rotational periods as the original one. Their relative shapes differ slightly, and importantly, their major power density  peaks can be at periods both longer and shorter than the underlying rotation period. This suggests  that not only these waves can cause differences between  the ``observed" lightcurve periodicity and the underlying rotation period, but also {\it the degree of this deviation} may vary with time.  }

{\btt For models with weaker frictional drags, the development of strong equatorial westward jets induce shifts in the period of variability. Panel e and f in Figure \ref{fig.periodogram} show periodograms for the case with $\tdrag=10^6$ and $10^7$ s, respectively. The first-order feature of them is that their major peaks shift to longer periods compared to the rotation period, and the shift is stronger in the case with $\tdrag=10^7 $ s. }

\section{Discussion and Conclusions}
\label{ch.conclusion}

Existence of large-scale zonally propagating waves at low latitudes in our simulations opens up an avenue to understand weather on isolated BDs and directly imaged EGPs, and its consequences on the observed lightcurve variability. Rapidly evolving isotropic storms and vortices are prevalent at mid-to-high latitudes, contributing to  lightcurve variability especially when the objects are viewed relatively pole-on (in which rotational modulation of lightcurve variability is diminished). Both dynamical features are driven by the cloud radiative feedback,  providing an essential physical mechanism to explain several types of time evolution of lightcurve variability (see a summary and analysis in \citealp{apai2017}). Eastward propagating Kelvin waves with phase speeds of a few hundred $\mps$ are sometimes dominant in our simulations, and the existence of these waves in atmospheres of BDs may explain the shorter rotation period of the atmosphere than that of the interior observed for a nearby BD  \citep{allers2020}. Our models predict that different equatorial waves that may travel in different zonal velocities could influence the lightcurves differently at different times. The interesting consequence is that these waves can cause differences between  the ``observed" lightcurve periodicity and the underlying rotation period, and {\it the degree of this deviation} may vary with time.   It will be interesting that the same observations performed by \cite{allers2020} could be repeated for the same system in the future to examine its long-term variability.  

Recently, \cite{vos2017}, \cite{vos2018} and \cite{vos2020} suggested  a viewing angle dependence of the observed near-IR colors and variability amplitude for a large sample of field BDs.  Our dynamical models provide  support for their observational results. Larger variability amplitude when viewed more equator-on is a natural result of the equatorial maximum of rotational variability along with our finding that  cloud patches reach maximum sizes at low latitudes (Figure \ref{fig.global_tq}, \ref{fig.global_flux}, \ref{fig.lightcurve} and \ref{fig.amplitude}). The vertical extent of zonal-mean cloud mixing ratio is higher at lower latitudes (Figure \ref{fig.tracer2zonalav}), which could be responsible for the redder near-IR colors when viewed more equator-on. The near-IR colors show a wide scatter in the color-magnitude diagram for mid-to-late L dwarfs (see color-magnitude diagrams for a large sample of BDs in, e.g., \citealp{faherty2016, liu2016}). Due to the latitudinal variation of the cloud  thickness at a given  rotation period and the dependence of global-average cloud thickness on varying rotation period, the different viewing angles and the variation of rotation periods of the field BDs might contribute to the scatter of observed near-IR colors. 
\cite{millar2020} showed that assuming two broad zonal bands with different cloud properties in each hemisphere, they can reproduce time-averaged polarization measured for the nearby BD Luhman 16B. Our models do not show clear zonally banded cloud structure like those in Jupiter and Saturn, but exhibit smooth equator-to-pole cloud thickness variations.  It is worthwhile to explore how can polarization  be produced by the smooth equator-to-pole cloud thickness variation with different viewing angles and  compare it to the measured   polarization.

Amplitudes of our simulated lightcurves typically range from 0.5 percent to a few percent depending on the rotation period and viewing angle, consistent with those found in the majority of  observed lightcurves \citep{buenzli2014,radigan2014,radigan2014b,wilson2014,metchev2015,vos2018}. We did not reproduce variability with amplitude much greater than a few percent. Yet, several field BDs and free-floating low-mass objects have shown large variability with peak-to-peak amplitudes greater than 10 percent (e.g., \citealp{buenzli2012,apai2013,buenzli2015b,lew2016,apai2017,zhou2020}). Either cloud patches with  sizes larger than those in our simulations or a greater horizontal contrast of outgoing thermal flux are required to explain those very large varability amplitudes. Our models only occupy a very limited parameter space. Further exploring parameter space, including broad range of surface gravity, atmospheric temperature and varying cloud particle size, will yield richer dynamical behaviors.  

Our models are highly idealized in the sense that radiative transfer and cloud formation (including the cloud microphysics and the treatment of sub-grid-scale structure) are vastly simplified in order to emphasize the role of atmospheric dynamics. {\btt Our modeled cloud structures capture the first-order behavior of cloud formation that has been shown by previous cloud formation models of BDs---a sharp cloud base typically emerges around the condensation level and the cloud  mixing ratio smoothly decreases with decreasing pressure due to mixing (see reviews by, e.g., \citealp{helling2013,marley2015,helling2019}). Our cloud scheme neglects temperature feedback on cloud formation, which could influence locations of the cloud base, and it certainly does not capture all the sophisticated microphysical processes.   Future endeavours should include better representation of radiative transfer, cloud microphysics and  parameterization of sub-grid cloud structure.}

Finally, we summarize our major findings in this study as follows:
\begin{itemize}

    \item Vigorous atmospheric circulation can be triggered and maintained by cloud radiative feedback in conditions appropriate for BDs and directly imaged EGPs.   {\btt When the bottom frictional drag is strong, zonal flows in the deep layers of our models are weak (with speeds much smaller than 100$\mps$).  In the observable layer where clouds form, } mid-to-high latitudes are dominated by isotropic vortices, with thick clouds forming in  anticyclones and thin clouds or clear sky in  cyclones. This is consistent with the results of previous $f$-plane models \citep{tan2020bd}. At low latitudes, large-scale zonally propagating waves with both eastward and westward phase speeds are the dominant dynamical feature. At a given rotation period, sizes of storms and vortices are typically larger at lower latitudes than those at higher latitudes. The overall sizes of storms are larger when the rotation period is longer.
    
    \item Lightcurves from our simulations have amplitudes from 0.5 percent to several percent, consistent with the majority of observed lightcurves. For a given rotation period, the lightcurve amplitude decreases with increasing viewing angle (0$^{\circ}$ means equator-on and 90$^{\circ}$ means pole-on), while it typically increases with increasing rotation period at a given viewing angle. When the bottom drag is strong,  zonally propagating waves at low latitudes have typical phase speeds of a few hundred  $\mps$. They can cause short-term evolution of lightcurves via wave beating effects, qualitatively similar to the observed lightcurves of some field BDs as summarized in \cite{apai2017}. The eastward Kelvin waves can cause the equatorial flux inhomogeneity  rotating faster than the underlying planetary rotation, which may help to explain the observed shorter rotation period of atmospheric features than that of the interior \citep{allers2020}. Isotropic storms and vortices at mid-to-high latitudes also contribute to the lightcurve variability.
    
    \item Clouds are generally mixed to higher altitudes near the equator than at high latitudes due to the stronger effect of rotation at high latitudes. This supports the observed redder IR colors for objects viewed more equator on \citep{vos2017,vos2020}.
    
    \item {\btt We expect that the strong-drag models might be appropriate for BDs and directly imaged EGPs because efficient convective mixing in the interior is expected to suppress strong zonal flows near the top of convective zone. But we still perform experiments with weaker drags to explore dynamics in the weak-drag regime. We find that robust zonal jets with speeds from several hundred to more than 2000$\mps$ can form in our weak-drag models, with typically a broad westward equatorial jet and a high-latitude eastward jet in each hemisphere. Similar to the weak-drag cases, vortices form at mid-to-high latitudes and equatorially trapped waves form at low latitudes. Both the zonal propagation of equatorial waves and the spacial distribution of vortices are affected by the presence of strong jets. Simulated lightcurves show longer periodicity than the underlying rotation period due to the strong westward equatorial jets.  The meridionally broad jet structure may  be related to the efficient potential vorticity mixing associated with intense eddies.}
    
    \item The origin of the equatorially propagating waves in our simulations is likely related to the self-excitation by cloud radiative feedbacks. Physical properties of the equatorially symmetric eastward waves resembles properties of both adiabatic free Kelvin waves and forced-damped waves triggered by a stationary equatorial heat source. Linear stability theory of waves coupled with cloud radiative feedback may help to explain the origin of our simulated waves but fails to explain  their propagation. 
\end{itemize}

\section*{Acknowledgements}
We thank Tad Komacek and Xi Zhang for comments on the draft and Ray Pierrehumbert for discussion. X.T. acknowledges support from the European community through the ERC advanced grant EXOCONDENSE (PI: R.T. Pierrehumbert). This work was completed with resources provided by the department of Physics at University of Oxford and the Lunar and Planetary Laboratory at University of Arizona. 

\section*{Data availability}
The data underlying this article will be shared on reasonable request to the corresponding author.



\bibliographystyle{mnras}
\bibliography{draft} 




\appendix

\section{Hovm$\ddot{\rm o}$ller diagrams at mid latitudes}
\label{ch.hovmoller}

\begin{figure*}
	\includegraphics[width=1.7\columnwidth]{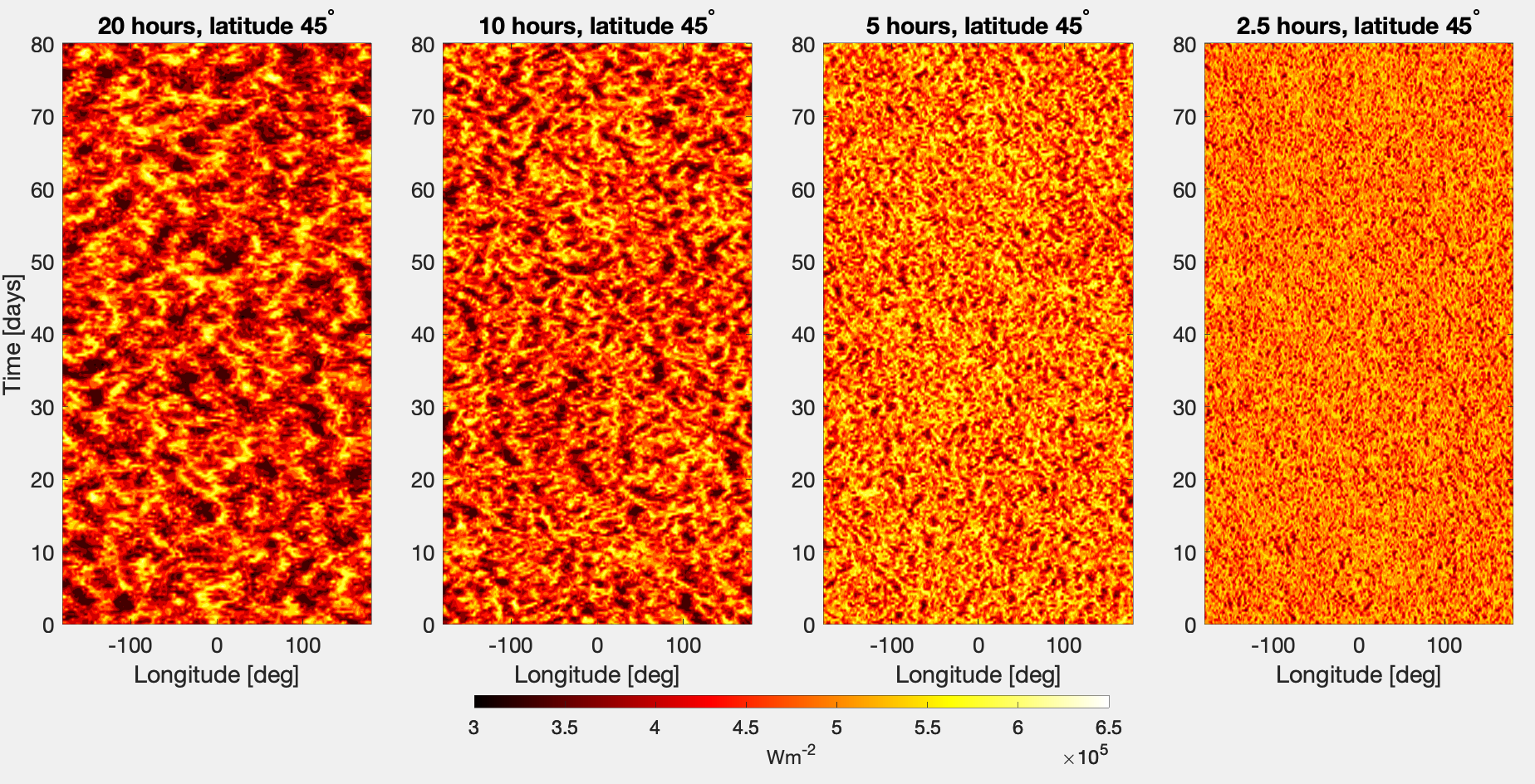}
    \caption{Hovm$\ddot{\rm o}$ller diagrams (longitude-time sequence) showing the time evolution of outgoing thermal flux at $45^{\circ}$ latitude  for models with four rotation periods of 20, 10, 5 and 2.5 hours (from the left to right) and  a  drag timescale $\tau_{\rm{drag}}=10^5$ s.   }
\label{fig.equatortime45}
\end{figure*}

In Figure \ref{fig.equatortime45}, we show the Hovm$\ddot{\rm o}$ller diagrams of outgoing thermal flux at $45^{\circ}$ latitude as a function of longitude and  time for models with different rotation period. These regions are dominated by stochastically evolving vortices, and there is no systematic eastward or westward propagation seen in these diagrams, which is in stark contrast to the equatorial disturbances shown in Figure \ref{fig.equatortime}. 

\section{Cloud radiatively induced equatorial waves}
\label{ch.theory}

Equatorially trapped waves in our simulations play  vital roles in the evolution of simulated lightcurves, and may be responsible for the puzzling time  evolution of some observed lightcurves \citep{apai2017}. Given their central importance, we explore their nature using a linear wave theory that is coupled to cloud radiative effect. We will show that the linear theory may be used to explain the energetic sources of our simulated waves, but cannot capture the propagating nature of the simulated waves.


The analysis of cloud radiatively induced instability was first carried out by \cite{gierasch1973} in the quasi-geostrophic system and inertia gravity wave ($f$-plane) system.  In the absence of rotation, 2D hydrostatic gravity waves have a set of pure unstable growing modes (no propagation) and sets of attenuating, eastward and westward propagating mode. In a $f-$plane approximation, inertia gravity waves with small horizontal wavelengths can host purely unstable modes but unstable modes  cease at larger horizontal scales. In a quasi-geostrophic flow, unstable modes are possible for axially symmetric flows and flows with nonzero $\beta$ \citep{gierasch1973}. We now extend this theory to the equatorial $\beta$ plane.

\subsection{Cloud radiative effect upon thermodynamics}
We first present the thermodynamics related to cloud radiative effect following \cite{gierasch1973}. Assuming that the change of outgoing thermal flux is due to brightness temperature deviation that is caused by either actual temperature variation or the cloud-top altitude variation, we can  write the change of emitted flux as 
\begin{equation}
\delta(\sigma T_c^4) = 4\sigma T_c^3(\delta T_c - \Gamma_c \delta z_c),
\label{eq.fluxchange}
\end{equation}
where $T_c$ is the brightness temperature, $\sigma$ is the Stefan-Boltzmann constant, $\delta z_c$ is the change of cloud-top altitude, and $\Gamma_c=|\frac{dT_c}{dz_c}$|. Because of the change of outgoing radiative flux, the atmospheric column is no longer in equilibrium and so net heating/cooling must occur. Denoting $Q$ as the heating rate per unit mass, $\rho$ the gas density and $D$ the cloud thickness, we have 
\begin{equation}
\int_0^D Q\rho dz=-\delta(\sigma T_c^4) = 4\sigma T_c^3(\Gamma_c \delta z_c - \delta T_c).
\end{equation}
For analytic simplicity,  $Q$ is assumed to obey the following linear relation at any level:
\begin{equation}
Q=\frac{4\sigma T_c^3}{M}(\Gamma_c \delta z - \delta T),
\label{eq.cloudinst3}
\end{equation}
where $M=\int_0^D \rho dz$. Moreover, clouds are assumed to be perfectly advected by the flow, i.e., sedimentation is negligible compared to the vertical advection, so that the rate of change in cloud-top altitude is simply the vertical velocity $w$. Taking the time derivative of Eq. (\ref{eq.cloudinst3}), we have
\begin{equation}
\frac{\partial Q}{\partial t} = \frac{4\sigma T_c^3}{M} \left(\Gamma_c w -  \frac{\partial T}{\partial t} \right).
\label{eq.heating}
\end{equation}

Then we assume a basic state of the atmosphere at rest with small horizontal temperature perturbations. The linearized equation of thermodynamics in height coordinates expressed using temperature is written:
\begin{equation}
\frac{\partial T}{\partial t} + w \left(\frac{\partial \overline{T}}{\partial z} + \frac{g}{c_p} \right) = \frac{Q}{c_p}.
\end{equation}
Combining with Eq. (\ref{eq.heating}) we have the linearized thermodynamic equation
\begin{equation}
\frac{\partial^2 T}{\partial t^2} + \frac{\partial w}{\partial t}\overline{\Gamma} = \frac{1}{\tau} \left(\Gamma_c w - \frac{\partial T}{\partial t} \right),
\label{eq.therm1}
\end{equation}
where $\overline{\Gamma} = \frac{d\overline{T}}{dz} +\frac{g}{c_p}$, and $\frac{1}{\tau} = \frac{4\sigma T_c^3}{c_p M}$. Here $\tau$ represents a radiative timescale of the cloud-forming atmosphere. 

By just the terms involving vertical velocity in Equation (\ref{eq.therm1}), $ \frac{\partial w}{\partial t}\overline{\Gamma} = \frac{1}{\tau}\Gamma_c w$, an instability is obvious with a growth rate of  $\lambda_c=\frac{\Gamma_c}{\overline{\Gamma \tau}}$. Now we consider the full linearized dynamical equations below.

\subsection{Rossby, mixed Rossby-gravity and inertia-gravity modes}
\label{ch.rossby}
We start with the linearized dynamical equations with a  basic state at rest in an equatorial $\beta$ plane, where the Coriolis parameter is written $f=\beta y$. The zonal and meridional angular momentum , hydrostatic balance, and continuity equations are, respectively
\begin{equation}
    \frac{\partial u}{\partial t} - \beta y v = - \frac{\partial \phi}{\partial x},
    \label{eq.u1}
\end{equation}
\begin{equation}
    \frac{\partial v}{\partial t}+\beta y u =- \frac{\partial \phi}{\partial y},
    \label{eq.v1}
\end{equation}
\begin{equation}
    \frac{\partial \phi }{\partial z}= \frac{gT}{T_0},
    \label{eq.hydrostatic}
\end{equation}
\begin{equation}
    \frac{\partial u}{\partial x}+\frac{\partial v}{\partial y}+\left(\frac{\partial}{\partial z}-\frac{1}{H}\right)\omega = 0,
    \label{eq.continuity1}
\end{equation}
where $T_0$ is a reference temperature, $z=-H\ln(p/p_0)$ is the vertical coordinate at log pressure, $H=RT_0/g$ is the scale height, and $\phi$ is the geopotential. Equations (\ref{eq.therm1}) to (\ref{eq.continuity1}) forms a closed set. 

We seek wave-like solutions:
\begin{equation}
     \begin{aligned}
    & \{u,v,\phi,T,\omega\} = \\
     &  \{\tilde{u}(y),\tilde{v}(y),\tilde{\phi}(y),\tilde{T}(y),\tilde{\omega}(y)\}
    e^{z/(2H)}\sin(\pi z/D)e^{ik_x x+\lambda t} ,
    \label{eq.wavesolution}
\end{aligned}
\end{equation}
where $\tilde{u}(y),\tilde{v}(y),\tilde{\phi}(y),\tilde{T}(y)$ and $\tilde{\omega}(y)$ are functions of $y$ only. To satisfy the form of equatorially trapped waves, the  boundary condition is typically applied to ensure that the disturbances vanish when $|y|\rightarrow \infty$.
Repeating the derivations for the classic equatorial waves (e.g., \citealp{matsuno1966,holton2012}), we arrive at an equation for $\tilde{v}(y)$:
\begin{equation}
    \left[\mathcal{L}^2_d\left(\frac{\lambda^2(\frac{1}{\tau}+\lambda)}{c^2_g(\lambda_c-\lambda)} - k^2_x + \frac{ik_x\beta}{\lambda}\right) -  Y^2 \right] \tilde{v} + \frac{d^2 \tilde{v}}{dY^2}=0,
    \label{eq.dispersion3}
\end{equation}
where $c^2_g=N^2/k^2_z$, $N^2=g\overline{\Gamma}/T_0$, $Y=y/\mathcal{L}_d$, and $\mathcal{L}_d$ is the equatorial deformation radius modified by the diabatic cloud radiative effect:
\begin{equation}
    \mathcal{L}^4_d \equiv \frac{c^2_g(\lambda - \lambda_c)}{(\frac{1}{\tau}+\lambda)\beta^2}.
\end{equation}
{\btt Note that in the equatorial theory of \cite{hayashi1970} that considered latent heating effects, a complex equatorial deformation was also possible.
In the adiabatic limit of $\tau\rightarrow\infty$, the deformation radius recovers the classic definition $\mathcal{L}_d=\sqrt{c_g/\beta}$ and the dispersion recovers the classic dispersion relation of adiabatic,  unforced equatorially trapped waves \citep{matsuno1966}. }
 In the limit of $\tau\rightarrow 0$, there is no solution that satisfies the boundary condition. 
The following relation has to be met for the given boundary condition:
\begin{equation}
    \mathcal{L}^2_d\left(\frac{\lambda^2(\frac{1}{\tau}+\lambda)}{c^2_g(\lambda_c-\lambda)} - k^2_x + \frac{ik_x\beta}{\lambda}\right) = 2n+1; \quad\quad n=0,1,2,...
    \label{eq.dispersion_final}
\end{equation}
Then, the solution has the form
\begin{equation}
    \tilde{v}(Y) = \tilde{v}_0{\rm H}_n(Y)\exp(-Y^2/2), 
\end{equation}
where $\tilde{v}_0$ is a constant with units of velocity, and ${\rm H}_n(Y)$ designates the $n$th {\it Hermite polynomial}.  In addition, solutions satisfy the boundary condition if $\exp(-Y^2/2)$ diminishes when $|y|\rightarrow \infty$, which requires that the real part of $\mathcal{L}_d^2$ is positive. 

\begin{figure}
	\includegraphics[width=1\columnwidth]{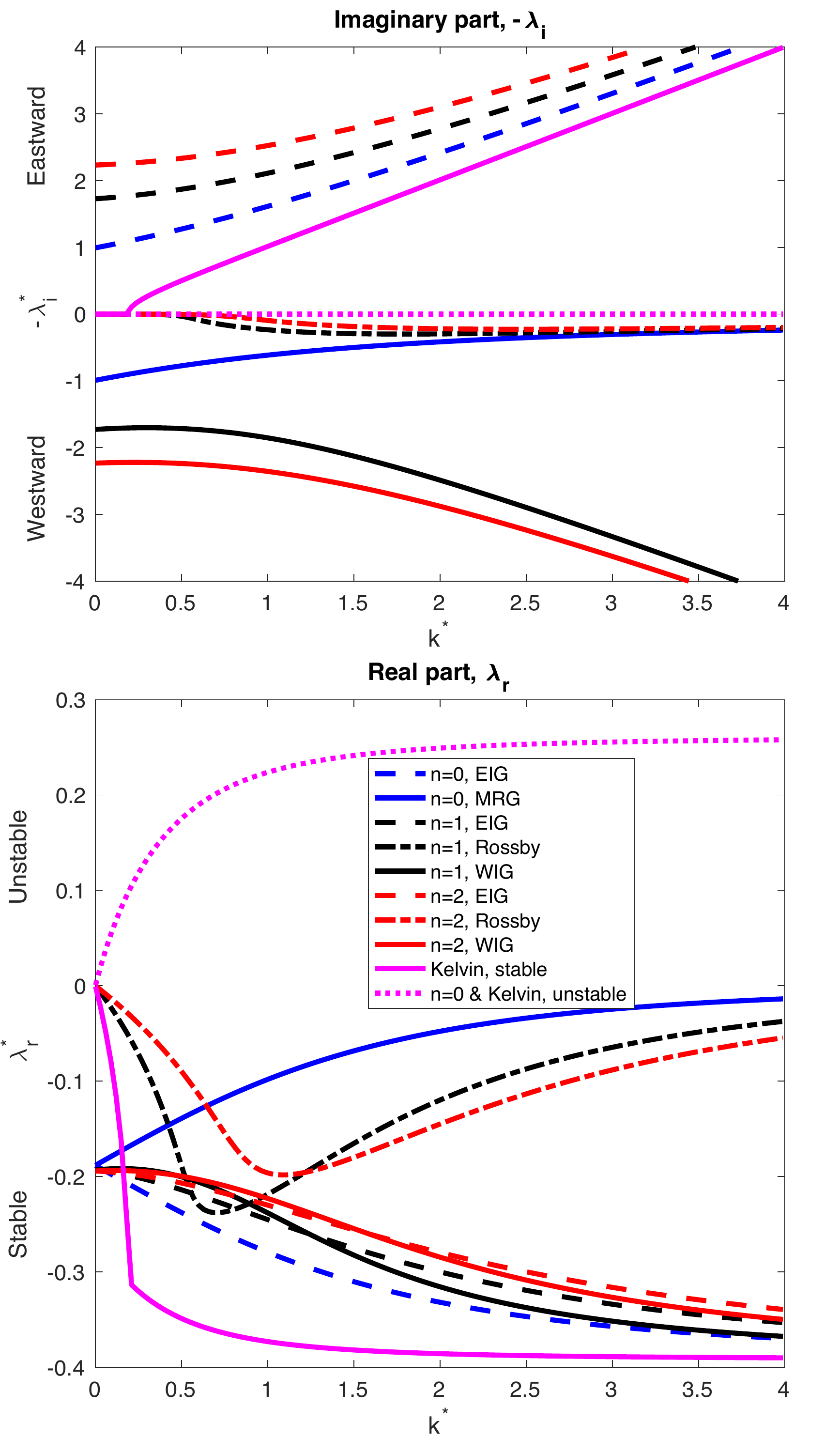}
    \caption{
    {\it Top panel}: dispersion relations of the imaginary components  $-\lambda_i$, i.e., the propagating components, for cloud-radiatively-coupled free equatorial waves. Frequency and zonal wavenumbers have been non-dimensionalized as $\lambda^{\ast}=\lambda/(\beta c_g)^{1/2}$ and $k^{\ast}=k(c_g/\beta)^{1/2}$. Different modes are labelled  in the right panel and with different colors and line styles. The dotted line corresponds to a special case in the Kelvin and $n=0$ unstable modes.  {\it Bottom panel}: dispersion relations of the real components  $\lambda_r$, i.e., the attenuating or growing components, for equatorial waves. Negative $\lambda_r$ represents attenuating modes and positive $\lambda_r$ represents growing modes.
    }
\label{fig.eqwaveanal}
\end{figure}

We first seek solutions with real, positive $\lambda$. Possible solutions should be in the range between 0 and $\lambda_c$. Outside this range, there is no solution that satisfies a real $k_x$, which is required to have wave-like zonal disturbances. 
Equating the imaginary part in the LHS of Equation (\ref{eq.dispersion_final}) to zero, one obtains 
\begin{equation}
    \lambda^3+\frac{1}{\tau}\lambda^2+ c^2_g k^2_x \lambda - \lambda_c c^2_g k^2_x = 0.
    \label{eq.realsolution1}
\end{equation}
There is one positive real root that satisfies $0<\lambda<\lambda_c$. This dispersion relation is that of the pure hydrostatic gravity waves, the same as Equation (28) in \cite{gierasch1973} when $f=0$. Equating the remaining real parts in Equation (\ref{eq.dispersion_final}), one obtains
\begin{equation}
    (2n+1)^2\left(\lambda^3+\frac{1}{\tau}\lambda^2\right)+ c^2_g k^2_x \lambda - \lambda_c c^2_g k^2_x = 0.
    \label{eq.realsolution2}
\end{equation}
Only when $n=0$ (the mixed Rossby gravity modes), both equations (\ref{eq.realsolution1}) and (\ref{eq.realsolution2}) can be simultaneously satisfied, and there is a set of purely unstable, non-propagating $n=0$ modes. In this case, $\mathcal{L}_d^2$ is purely imaginary and fails the strict meridional boundary condition. Nevertheless, by solving equation (\ref{eq.dispersion_final}) numerically, we cannot find other modes that contain a positive real part of $\lambda$. All other modes  have both  negative real parts and  imaginary parts, and they resemble the classic equatorial waves---the propagating Rossby, MRG and inertia-gravity modes but with damping in the wave amplitudes due to thermal radiation. The dispersion relations for all possible solutions are plotted as solid curves in Figure \ref{fig.eqwaveanal}, in which the top panel shows the $-\lambda_i$ (in the same format as that plotted in classic literature of equatorial waves, e.g., \citealp{holton2012}) and the bottom panel shows the $\lambda_r$, where $\lambda=\lambda_r+i \lambda_i$. Despite that the special non-propagating unstable mode in Equation (\ref{eq.realsolution1}) does not strictly satisfy the lateral boundary condition, at least their disturbances do not amplify with  $|y|\rightarrow\infty$. So we still plot the dispersion relation as the dotted lines in Figure \ref{fig.eqwaveanal}.

\subsection{Kelvin modes}
\label{ch.kelvin}
The   Kelvin mode is a special mode in which the meridional velocity is zero. The governing equations for the zonal and meridional angular momentum, and continuity  are simplified to
\begin{equation}
    \frac{\partial u}{\partial t} = -\frac{\partial \phi}{\partial x},
    \label{eq.kelvin.u}
\end{equation}
\begin{equation}
    \beta y u=-\frac{\partial\phi}{\partial y},
    \label{eq.kelvin.v}
\end{equation}
\begin{equation}
    \frac{\partial u}{\partial x}+\left(\frac{\partial}{\partial z}-\frac{1}{H}\right) \omega=0.
    \label{eq.kelvin.cont}
\end{equation}
We make use of equations (\ref{eq.kelvin.u}), (\ref{eq.hydrostatic}), (\ref{eq.kelvin.cont}) and (\ref{eq.therm1}) and assume wave-like solutions of equation (\ref{eq.wavesolution}), then we  obtain the following dispersion relation
\begin{equation}
    \lambda^3+\frac{1}{\tau}\lambda^2+ c^2_g k^2_x \lambda - \lambda_c c^2_g k^2_x = 0.
\end{equation}
This is the same as Equation (\ref{eq.realsolution1}) for $n=0$ modes, and growing modes with positive $\lambda$ exist.  Additional constraints from the lateral boundary condition should be satisfied. Combining equations (\ref{eq.kelvin.u}) and (\ref{eq.kelvin.v}), one obtains
\begin{equation}
    \tilde{u}=\tilde{u}_0 \exp\left(\beta\frac{k_x\lambda_i+ik_x\lambda_r}{2(\lambda_i^2+\lambda_r^2)}  y^2\right ),
\end{equation}
where $\tilde{u}_0$ is the amplitude of the perturbation zonal velocity at the equator, and $\lambda$ is written as $\lambda=\lambda_r+i\lambda_i$.
If one restricts $\tilde{u}$ to vanish with $|y|\rightarrow \infty$,  $k_x\lambda_i<0$ needs to be satisfied. With this regard, the purely unstable modes fail to satisfy the boundary condition, and only the eastward propagating but decaying modes are valid solutions. This reaches the same conclusion as the $n=0$ unstable modes. The dispersion relation of the unstable  Kelvin modes are also represented as dotted lines in Figure \ref{fig.eqwaveanal}.

Our key finding is that there is only one set of growing but non-propagating modes corresponding to the Kelvin and $n=0$ modes that may be marginally relevant. This may provide a way to excite the Kelvin waves, $n=0$ MRG waves and $n=0$ eastward inertia gravity waves seen in our simulations.  Other propagating modes have properties quantitatively similar to the adiabatic free modes but with damping of their amplitudes due to thermal radiation.


\bsp	
\label{lastpage}
\end{document}